\documentclass[a4paper,12pt]{article}
\pdfoutput=1
\usepackage[utf8]{inputenc}
\usepackage{enumerate}
\usepackage{jheppub,undertilde}
\usepackage{feynmp-auto}
\usepackage{tikz}
\usetikzlibrary{matrix,positioning,decorations.pathreplacing}
\usepackage{graphicx}
\usepackage{color}
\usepackage{float}
\usepackage{soul}
\usepackage{siunitx}
\usepackage{arydshln}
\usepackage{bbold}
\usepackage{verbatim}
\usepackage{hyperref}
\usepackage{tikz}
\usepackage{slashed}
\usepackage{lineno}

\usetikzlibrary{arrows,positioning} 
\tikzset{
    >=stealth',
    punkt/.style={
           rectangle,
           rounded corners,
           draw=black, very thick,
           text width=6.5em,
           minimum height=2em,
           text centered},
    pil/.style={
           ->,
           thick,
           shorten <=2pt,
           shorten >=2pt,}
}

\makeatletter
\def\user@resume{resume}
\def\user@intermezzo{intermezzo}
\newcounter{previousequation}
\newcounter{lastsubequation}
\newcounter{savedparentequation}
\renewenvironment{subequations}[1][]{%
      \def\user@decides{#1}%
      \setcounter{previousequation}{\value{equation}}%
      \ifx\user@decides\user@resume 
           \setcounter{equation}{\value{savedparentequation}}%
      \else  
      \ifx\user@decides\user@intermezzo
           \refstepcounter{equation}%
      \else
           \setcounter{lastsubequation}{0}%
           \refstepcounter{equation}%
      \fi\fi
      \protected@edef\theHparentequation{%
          \@ifundefined {theHequation}\theequation \theHequation}%
      \protected@edef\theparentequation{\theequation}%
      \setcounter{parentequation}{\value{equation}}%
      \ifx\user@decides\user@resume 
           \setcounter{equation}{\value{lastsubequation}}%
         \else
           \setcounter{equation}{0}%
      \fi
      \def\theequation  {\theparentequation  \alph{equation}}%
      \def\theHequation {\theHparentequation \alph{equation}}%
      \ignorespaces
}{%
  \ifx\user@decides\user@resume
       \setcounter{lastsubequation}{\value{equation}}%
       \setcounter{equation}{\value{previousequation}}%
  \else
  \ifx\user@decides\user@intermezzo
       \setcounter{equation}{\value{parentequation}}%
  \else
       \setcounter{lastsubequation}{\value{equation}}%
       \setcounter{savedparentequation}{\value{parentequation}}%
       \setcounter{equation}{\value{parentequation}}%
  \fi\fi
  \ignorespacesafterend
}
\makeatother

\newcolumntype{C}{>{$}c<{$}}

\newcommand{\SM}{\text{SM}}
\newcommand{\SO}{\text{SO}}
\newcommand{\SU}{\text{SU}}
\newcommand{\U}{\text{U}}

\newcommand{\TeV}{\text{TeV}}
\newcommand{\GG}{\mathcal{G}}
\newcommand{\HH}{\mathcal{H}}

\newcommand{\OO}{\mathcal{O}}

\newcommand{\tr}{\text{Tr}}

\definecolor{darkred}{rgb}{0.8,0.2,0.3}

\definecolor{darkgreen}{rgb}{0.2,0.8,0.3}

\newcommand{\appropto}{\mathrel{\vcenter{
  \offinterlineskip\halign{\hfil$##$\cr
    \propto\cr\noalign{\kern2pt}\sim\cr\noalign{\kern-2pt}}}}}

\usepackage{cleveref}
\allowdisplaybreaks
\preprint{SISSA 16/2019/FISI}

\title{Composite 2HDM with singlets: a viable dark matter scenario}

\author[a,b]{Alessandro Davoli,}
\author[a,b]{Andrea De Simone,}
\author[b]{David Marzocca,}
\author[a,b]{Alessandro Morandini}

\affiliation[a]{SISSA, via Bonomea 265, 34136 Trieste, Italy}
\affiliation[b]{INFN Sezione di Trieste, via Bonomea 265, 34136 Trieste, Italy}

\emailAdd{alessandro.davoli@sissa.it}
\emailAdd{andrea.desimone@sissa.it}
\emailAdd{david.marzocca@sissa.it}
\emailAdd{alessandro.morandini@sissa.it}

\abstract{
We study the non-minimal composite Higgs model with global symmetry $\text{SO(7)}$ broken to $\text{SO(5)}\times \text{SO(2)}$. The model results in a composite Two-Higgs doublet model (2HDM) equipped with two extra singlets, the lightest of which can be a viable dark matter candidate. The model is able to reproduce the correct dark matter relic density both via the usual thermal freeze-out and through late time decay of the heavier singlet. In the case of thermal freeze-out, it is possible to evade current experimental constraints even with the minimum fine tuning allowed by electroweak precision tests.}

\begin{document}
\maketitle

\section{Introduction}
\label{sec:intro}

The spectacular success of the Standard Model (SM) in describing
particle physics phenomena, 
culminated with the discovery of the Higgs boson
\cite{Aad:2012tfa,Chatrchyan:2012xdj}, 
still leaves us with several open problems.
Two of the most pressing questions to address are:
what is the dynamics protecting the electroweak (EW) scale
from large ultraviolet (UV) corrections?
What is the nature of the dark matter (DM) of the universe?

The  Composite Higgs (CH) paradigm  \cite{Kaplan:1983fs,
Georgi:1984af,Kaplan:1983sm,Dugan:1984hq} provides a very appealing framework to answer both questions at once, see e.g. the reviews \cite{Contino:2010rs,Bellazzini:2014yua,Panico:2015jxa}.
In CH models,
a new strongly-coupled sector symmetric under a global symmetry $\mathcal{G}$ 
is assumed to exist above the electroweak scale.
The Higgs boson arises
as a pseudo-Nambu-Goldstone boson (pNGB) of the spontaneous breaking
 $\mathcal{G}\to \mathcal{H}$ at a scale $f$. The Higgs boson mass is naturally light, 
originating from explicit breaking of $\mathcal{G}$, in a similar fashion to
what happens for the QCD pion.
In most Composite Higgs constructions other particles in addition to the Higgs doublet arise from the symmetry breaking, depending on the specific breaking pattern.
In some cases, one of these may be stable and thus possibly have the right properties
to account for the non-baryonic DM component of the universe.
Such a scenario is one of the few in which the DM candidate is naturally at the same mass scale as the Higgs boson since the DM explanation is tightly linked with the solution of the  hierarchy problem (the other notable example being neutralinos in TeV-scale supersymmetry).

From the point of view of low-energy model building, the first step is to choose the
symmetry breaking pattern $\mathcal{G}$ and $\mathcal{H}$. The most minimal CH model including a DM candidate among the pNGBs is based on the $\SO(6)/\SO(5)$ coset. This scenario includes a real singlet in addition to the Higgs doublet and its DM phenomenology has been widely studied in the literature (see e.g. refs.~\cite{Frigerio:2012uc,Marzocca:2014msa,Fonseca:2015gva,Bruggisser:2016ixa,Gripaios:2009pe}).
Other studies focused on several different symmetry-breaking patterns, including for example $\SO(6)\to \SO(4)\times\SO(2)$ \cite{Fonseca:2015gva}, $\SO(7)\to G_2$ \cite{Ballesteros:2017xeg}, $\SO(7)\to\SO(6)$ \cite{Balkin:2017aep,Balkin:2018tma,DaRold:2019ccj}, $\SO(7) \to \SO(5)$ \cite{Chala:2018qdf},  $\SU(4)\times\SU(4)\to\SU(4)$ \cite{Ma:2015gra,Ma:2017vzm}, $\SU(5)\to\SO(5)$ \cite{Balkin:2017yns}, and $\SU(6)\to\SO(6)$ \cite{Cacciapaglia:2019ixa} (see also \cite{Alanne:2018xli,Carmona:2015haa}).
The motivation for studying one particular coset can arise either from an underlying UV completion, or from peculiarities in pNGB field content or in their dynamics which make the phenomenology of the model interesting to explore. In this paper, we follow the latter guideline.

We construct and study a CH model based on $\SO(7)\to \SO(5)\times\SO(2)$.
The pNGB field content of this theory consists of two Higgs doublets and two real scalars, the lightest of which is stable and is our dark matter candidate.
As will be described in more detail below and in the rest of the paper, the presence of the second doublet will be important to relax the constraints from electroweak precision tests (EWPTs) and DD, while the second singlet can offer an interesting alternative mechanism for DM production in the early Universe.
Another desirable feature of this coset is the absence of a Wess-Zumino-Witten anomaly. This model is an extension of the composite Two-Higgs doublet model (2HDM) based on the coset $\SO(6) \to \SO(4) \times \SO(2)$, studied in detail in refs.~\cite{Mrazek:2011iu,Fonseca:2015gva,DeCurtis:2018iqd,DeCurtis:2018zvh}. Several features of our setup are shared with this smaller coset, which however does not include the two singlets pNGB, the lightest of which is our DM candidate.

In this paper, we build the low-energy effective theory of the pNGBs using the tools of Callan-Coleman-Wess-Zumino (CCWZ) construction \cite{Coleman:1969sm, Callan:1969sn} and naive dimensional analysis (NDA) \cite{Manohar:1983md,Panico:2011pw},
paying particular attention to the spectrum and interactions of the pNGBs.

A viable DM candidate needs to satisfy a variety of phenomenological 
constraints. Most notably, it should reproduce the correct relic abundance and it should not be excluded by direct detection experiments.
This implies severe constraints on the model parameters
and, in the context of CH models, typically 
requires $f$ much larger than the electroweak scale, 
and then relatively large fine tuning.
Instead, we find that a viable DM candidate consistent with all phenomenological constraints can be achieved in this model, without paying the price of an excessive fine tuning on the symmetry breaking scale $f$ with respect
to the one dictated by EWPTs. This is largely due to the contribution of the second Higgs doublet, which helps both to partially compensate the SM Higgs contribution to direct detection of DM, as well as to relax the EWPT constraints.

Furthermore, we find that DM production may be different from the usual thermal freeze-out mechanism, and  proceed non-thermally through decays of the heavier singlet pNGB.
This feature is possible because of the richer structure of the model, and (to the best of our knowledge) it is novel in DM models within the CH paradigm.

The present paper is organized as follows: in~\cref{sec:model}, we introduce and construct the effective Lagrangian of the model, while in~\cref{sec:dynamics} we describe the spectrum and the interactions of the pNGBs; in~\cref{sec:ThDM}, we study the thermal DM candidate of the model, how it can achieve the correct relic abundance, and the corresponding constraints from LHC searches, direct detection (DD) and indirect detection (ID); in~\cref{sec:non_thermal_DM}, the non-thermal production of DM by decays of the heavier singlet is studied; finally, we conclude in~\cref{sec:Conclusions}. The appendices contain technical supplementary material, such as the group generators (\cref{app:generators}), the constraints from EWPTs (\cref{app:EWPTHiggs}), the detailed expressions of the effective couplings of the NGBs interactions (\cref{app:effective_coupl}), and the calculation of the DM relic density (\cref{app:relicdensity}).

\section{Effective Lagrangian construction}
\label{sec:model}

In this section, we construct the low-energy effective theory, valid below the compositeness energy scale $\Lambda$, of pNGBs based on the coset $\SO(7)/\SO(5)\times \SO(2)$.
In the following three subsections, we present the pNGB fields of this theory, the details of the partial compositeness mechanism employed and the radiatively generated pNGB potential.

\subsection{Coset and the pseudo-NGB}
\label{sec:coset}

We consider a new strongly coupled sector lying at an energy scale $\Lambda = m_* \sim$ (few)~TeV and assume that it respects a global symmetry $\GG = \SO(7)$, spontaneously broken to a subgroup $\HH = \SO(5) \times \SO(2)$ at a scale $f \sim m_* / g_*$ by a condensate of the strong dynamics, where by $g_*$ we indicate a typical strong coupling of the composite sector.
This spontaneous symmetry-breaking pattern produces a set of ten NGBs transforming as a $({\bf 5}, {\bf 2})$ of $\HH$.

The global symmetry $\GG$ of the strong sector is explicitly broken by the SM gauging and the interactions which generate the Yukawa couplings.
This breaking induces a potential for the pNGB.
The EW gauge group $\GG_{\rm EW} = \SU(2)_L \times \U(1)_Y$ is assumed to be embedded in a subgroup $\HH'$, which in general does not coincide with the subgroup individuated by the vacuum of the theory, $\HH$.
This well known mechanism of vacuum misalignment between these two subgroups of $\GG$, shown schematically in~\cref{fig:gauge}, is responsible for the spontaneous breaking of $\GG_{\rm EW}$.

In general, we can then consider two basis of generators: one, $\{T_\theta\}$, related to the breaking $\mathcal{G}\to \mathcal{H}$, and a second one, $\{T\}$, related to $\mathcal{G}\to \mathcal{H}'$. The groups $\mathcal H$ and $\mathcal H'$ are misaligned by an angle $\theta$, which in general is a vector, since more than one field can acquire a vacuum expectation value (VEV). The corresponding vacua are related by a rotation matrix $r_\theta$ and, if we assume that the generators are normalized as $\tr\{T^AT^B\}=\delta_{AB}$, we have:
\begin{equation}
T_\theta = r_\theta\, T\,r_\theta^{-1}\,.
\end{equation}
We then introduce the Goldstone matrices in the two basis as:
\begin{equation}
U \equiv U(\Pi) = e^{i\frac{\sqrt2}{f} \Pi}\quad,\quad U_\theta \equiv U(\Pi_\theta) = e^{i\frac{\sqrt2}{f} \Pi_\theta}\,,
\end{equation}
where we defined $\Pi \equiv \Pi^I\hat T^I$ and $\Pi_{\theta} \equiv \Pi^I\hat T^I_{\theta}$, with $\hat T^I, \hat T^I_{\theta}$ being the broken generators in the two basis and $\Pi^I$ the respective pNGB fields.

The rotation $r_\theta$ can be obtained by considering the Goldstone matrix in the \emph{gauge} (non-rotated) basis $\{T\}$, and setting the NGBs at the corresponding VEVs, i.e.:
\begin{equation}
r_\theta \equiv U(\langle\Pi\rangle)\,.
\label{eq:def_r_theta}
\end{equation}
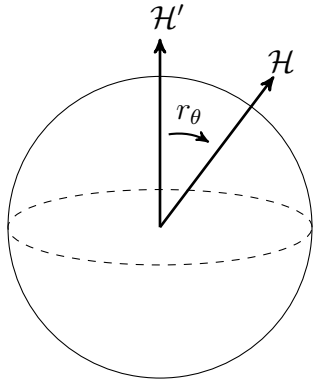
\begin{figure}
\begin{center}
\begin{tikzpicture}
\draw (0,0) circle (2cm);
\draw [dashed] (0,0) ellipse (2cm and .5cm);
\draw[->,line width=1pt] (0,0)--(0,2.5);
\draw[->,line width=1pt] (0,0)--(1.5,2);
\node at (-0.1,1.2) (a) {};
\node at (.85,1) (b) {}
edge[pil,<-, bend right=20] node[above] {$r_\theta$} (a);
\node at (0.1,2.8) {$\mathcal H'$};
\node at (1.6,2.2) {$\mathcal H$};
\end{tikzpicture}
\end{center}
\caption{\label{fig:gauge}The effect of gauging $\mathcal{H}'\subset \mathcal{G}$ is a breaking of the global symmetry due to vacuum misalignment.}
\end{figure}
Our choice for the generator basis is described as follows:
\begin{equation}
\{T_L,T_R,T_5,T_2,\hat T_1,\hat T_2\}\sim
\begin{tikzpicture}[baseline={([yshift=-.5ex]current bounding box.center)}, style1/.style={matrix of math nodes, every node/.append style={text width=#1,align=center,minimum height=#1}, nodes in empty cells, left delimiter=(, right delimiter=),}]
\matrix[style1=0.45cm](mat)
{
  & & & & & &  \\
  & & & & & & \\
  & & & & & & \\
  & & & & & & \\
  & & & & & & \\
  & & & & & & \\
  & & & & & & \\
};
\draw[dashed](mat-1-4.north east)--(mat-5-4.south east);
\draw[dashed](mat-4-1.south west)--(mat-4-5.south east);
\draw(mat-1-5.north east)--(mat-7-5.south east);
\draw(mat-5-1.south west)--(mat-5-7.south east);
\node at (mat-2-2.south east){$T_{L,R}$};
\node at (mat-2-5.south){$T_5$};
\node at (mat-5-2.east){$T_5$};
\node at (mat-5-5){0};
\node at (mat-6-6.south east){$T_2$};
\node at (mat-3-6){$\hat T_1$};
\node at (mat-3-7){$\hat T_2$};
\node at (mat-6-3){$\hat T_1$};
\node at (mat-7-3){$\hat T_2$};
\end{tikzpicture}\,.
\end{equation}
The generators of SO(5)$'\times$SO(2)$'$, whose expressions can be found in~\cref{app:generators}, are then block-diagonal in the basis we adopted. We indicate with the first letters of the alphabet the indices of a generic SO(7) transformation, $a,b,c,\dots = 1,\dots,7$. Because of its block-diagonal form, instead, an SO(5)$'\times$SO(2)$'$ transformation will have $\bar a,\bar b,\dots$ indices, where $\bar a=\{i,\mu\}$, with $i$ and $\mu$ being SO(5)$'$ five-plet and SO(2)$'$ doublet indices, respectively.
In the following, we do not distinguish between upper or lower indices, identifying however the first and second indices as row and column ones, respectively.

As already stated, the generators in the \emph{physical} vacuum ($\mathcal{G}/\mathcal{H}$) basis are related to these by $T_\theta=r_\theta\,T\,r_\theta^{-1}$; the virtue of this approach is that we can expand $U_\theta$ in the fields to extract the interactions, while keeping the exact, trigonometric, expression for the parameters related to the vacuum.

In order to identify the accidental symmetries of the theory and the quantum numbers of the pNGBs, it is useful to start by considering the limit of no misalignment, i.e. $r_\theta = {\bf 1}$. In this limit, and with the generator basis specified above, the NGB matrix takes the simple form:
\begin{equation}
\Pi= \hat T^I\Pi^I\equiv-\frac{i}{\sqrt2}
\begin{pmatrix}
\mathbb0_{5\times5} & \Phi_1 & \Phi_2 \\
-\Phi_1^T & 0 & 0 \\
-\Phi_2^T & 0 & 0
\end{pmatrix}~.
\end{equation}
The two NGB five-plets of $\SO(5)^\prime$ can be decomposed under representations of the custodial symmetry $\SO(4)_c \subset \SO(5)^\prime$ as $(\boldsymbol{5}, \boldsymbol{2}) = 2\times\boldsymbol{4} + 2\times\boldsymbol{1}$:
\begin{equation}
    \Phi_1=(\phi_1,\eta)^T~, \qquad
    \Phi_2=(\phi_2,\kappa)^T~. \qquad
    \label{eq:PNGB5plets}
\end{equation}
The two ${\bf 4}$ describe the two Higgs doublets of the theory: 
\begin{equation}
\phi_1 = \begin{pmatrix}
    G_1\\
    G_2\\
    G_3\\
    h
\end{pmatrix},\qquad
\phi_2 = \begin{pmatrix}
    \frac{-i}{\sqrt{2}}(H_+-H_-)\\
    \frac{1}{\sqrt{2}} (H_++H_-)\\
    H_0\\
    A_0
\end{pmatrix},
\end{equation}
where $h$ is the SM-like Higgs, $G_i$ are the would-be longitudinal polarizations of the EW gauge bosons, and $H_0$ and $A_0$ are the $CP$-even and -odd components of the neutral scalar, respectively. The lightest of the two singlets, $\eta$, will be the DM candidate.

As discussed in detail in ref.~\cite{Mrazek:2011iu}, it is useful to introduce a discrete transformation of the pNGBs, namely:
\begin{equation}
C_2=\text{diag}(1,1,1,1,1,1,-1)\,,
\end{equation}
acting as $U\to C_2UC_2$ on the Goldstone bosons matrix, under which $(\Phi_1,\Phi_2)\to(\Phi_1,-\Phi_2)$.
This symmetry is important to protect the second Higgs doublet from taking a sizeable VEV, thereby spontaneously violating custodial symmetry beyond the allowed limit.
As we will see, the interactions of the strong sector with SM fermions in general break this symmetry.
Another useful parity is:
\begin{equation}
P_7=\text{diag}(1,1,1,1,-1,1,1)~,
\end{equation}
under which $(\eta, \kappa) \to - (\eta, \kappa)$, which can thus stabilize the singlets. In the following, we impose $P_7$ as a symmetry of the theory.

An appealing property of the coset under consideration is that no Wess-Zumino-Witten term \cite{Wess:1971yu,Witten:1983tw} is generated since the fifth de Rham cohomology group of $\SO(7)/\SO(5)\times\SO(2)$ vanishes \cite{Davighi:2018inx,Davighi:2018xwn}\footnote{We thank Joe Davighi for sharing this result with us.}. This implies that the $P_7$ parity remains unbroken by the strong dynamics to all orders in the chiral expansion.
The parities of the NGBs and their representation under $\SO(4)^\prime$ are summarized in~\cref{tab:NGBreps}.

\begin{table}[t]
\begin{center}
\begin{tabular}{C|C|C|C}
\text{Field} & \text{SO(4)$'$} & C_2 & P_7\\
\hline
\phi_1&\mathbf{4}&+&+\\
\phi_2&\mathbf{4}&-&+\\
\eta&\mathbf{1}&+&-\\
\kappa&\mathbf{1}&-&-\\
\end{tabular}
\caption{\label{tab:NGBreps} Representations and quantum numbers of the NGB fields under SO(4)$'$, $C_2$ and $P_7$. $\phi_1$ can be identified as the SM Higgs doublet if $C_2$ is a symmetry of the NGB potential or more in general if $\phi_2$ does not take a VEV.}
\end{center}
\end{table}

Lagrangian terms which break explicitly the global symmetry $\mathcal G$ could in general also break these parities. In the following we show how, in general, $C_2$ is indeed  broken by the interaction of the top quark with the composite sector and by the potential, while $P_7$ can remain a good symmetry without being broken at any level. 

Compatibly with $P_7$ and $CP$, the general misalignment of the vacuum with respect to the gauged subgroup $\mathcal{H}^\prime$ can be described by two angles $\theta_1=\langle h \rangle/f$ and $\theta_2=\langle H_0 \rangle/f$.
The misalignment matrix is thus identified as:
\begin{equation}
    r_\theta = \left( \begin{array}{ccccccc}
         1 & 0 & 0 & 0 & 0 & 0 & 0  \\
         0 & 1 & 0 & 0 & 0 & 0 & 0  \\
         0 & 0 & c_{\theta_2} & 0 & 0 & 0 & s_{\theta_2}  \\
         0 & 0 & 0 & c_{\theta_1} & 0 & s_{\theta_1} & 0  \\
         0 & 0 & 0 & 0 & 1 & 0 & 0  \\
         0 & 0 & 0 & -s_{\theta_1} & 0 & c_{\theta_1} & 0  \\
         0 & 0 & -s_{\theta_2} & 0 & 0 & 0 & c_{\theta_2}
    \end{array}\right)~,
    \label{eq:rtheta}
\end{equation}
where $s(c)_{\theta_{1,2}} = \sin (\cos) \theta_{1,2}$.

\subsection{CCWZ Lagrangian}
\label{eq:pNGBLagr}

The leading operator describing the low-energy NGB effective theory is, in CCWZ language \cite{Coleman:1969sm, Callan:1969sn}:
\begin{equation}\label{eq:LCCWZ}
\mathcal{L}_{\Pi}^{(2)} \equiv \frac{f^2}{4}\tr\left[d^{(\theta)}_\mu d^{(\theta)\mu}\right]\,,
\end{equation}
where $d_\mu^{(\theta)} \equiv  i \sum_I \tr[U_\theta^{-1} D_\mu U_\theta \hat T_\theta^I] \hat T_\theta^I$.
This Lagrangian contains the kinetic terms for the Nambu-Goldstone fields, mass terms for the SM EW gauge bosons and their interactions with pNGBs, as well as an infinite series of two-derivative interactions among NGBs.
The mass of the $W$ boson is given by:
\begin{equation}
m_W^2 = \frac{g^2f^2}{4}(\sin^2\theta_1+\sin^2\theta_2)\,.
\end{equation}
It is then convenient to define:
\begin{equation}
\begin{split}
\sin \theta_1 &\equiv \sqrt\xi\cos\beta\,,\\
\sin \theta_2 &\equiv \sqrt\xi\sin\beta\,,
\end{split}
\label{eq:defXiBeta}
\end{equation}
so that, defining $m_W^2 = g^2v^2 / 4$, we get:
\begin{equation}
\xi \equiv \frac{v^2}{f^2} = \sin^2\theta_1+\sin^2\theta_2~, \qquad
\tan \beta = \frac{\sin \theta_2}{\sin \theta_1}\,,
\end{equation}
with $v=\SI{246}{\giga\electronvolt}$ being the SM VEV.

On the other hand, the prediction for the $Z$ mass is:
\begin{equation}
m_Z^2 = \frac{v^2(g^2+{g'}^2)}{4}\left[1-\frac{\xi}{4}\left(1-\cos4\beta\right)\right]\,,
\end{equation}
leading to a tree-level positive contribution to the $\hat T$ parameter:
\begin{equation}
(\Delta \hat T)_{\rm 2HDM} \approx \frac{\xi}{4} (1 - \cos 4\beta) \approx 2 \xi \beta^2 + \mathcal{O}(\xi \beta^4)~.
\label{eq:T_parameter}
\end{equation}
This is due to the fact that $\langle H_0 \rangle$ explicitly breaks $\SO(4)_c$ to $\SO(2)$. The different $CP$ structure with respect to \cite{Mrazek:2011iu} implies that we have custodial breaking even if we assume $CP$ to be preserved.
While $\mathcal{O}(1)$ values of $\beta$ are disfavored by EWPTs, as they would require very high fine tuning, values $\beta \lesssim 0.1$ are allowed. Such a contribution might even help to improve the fit to electroweak precision observables (see~\cref{app:EWPTHiggs} for more details). As shown in~\cref{sec:vacuum}, small values of $\beta$ are obtained naturally in this model.
Interestingly enough, the positive contribution to $\hat T$ given by~\cref{eq:T_parameter} can help to relax the usual EWPT limits on $\xi$. We find that for $\beta \approx 0.1$, a fine tuning up to $\xi \approx 0.08$ is compatible with both EW and Higgs data (see~\cref{app:EWPTHiggs}).

\subsection{Partial compositeness}
\label{sec:partial compositeness}

To couple the SM fermions to the Higgs field and generate their masses, we resort to the partial compositeness paradigm: the basic idea is that quarks are linearly coupled to fermionic operators $\mathcal O_{L,R}$ belonging to the strong sector \cite{Kaplan:1991dc}. In the following, we assume for simplicity that the operators coupled to the top quarks transform in the fundamental representation of SO(7), although other choices are possible (see e.g. \cite{Balkin:2018tma} for other representations of SO(7)).

As usual in Composite Higgs models, the group $\mathcal{G}$ has to be enlarged to correctly reproduce the SM quantum numbers: to this purpose we consider SO(7)$\times$U(1)$_X$, where the charge $X$ is $X=2/3$ for the top quark. The hypercharge is then identified with $Y=T_R^3+X$.

The $\mathbf7$ decomposes under SO(5)$'\times$SO(2)$'$ and SU(2)$_L\times$U(1)$_{Y}$ as:
\begin{equation}
\mathbf7_\frac{2}{3} = \left(\mathbf5,\mathbf1\right)_\frac{2}{3} \oplus (\mathbf1,\mathbf2)_\frac{2}{3}= \mathbf2_\frac{7}{6} \oplus \mathbf2_\frac{1}{6} \oplus \mathbf1_\frac{2}{3} \oplus \mathbf1_\frac{2}{3} \oplus \mathbf1_\frac{2}{3}\,.
\label{eq:decomposition_7}
\end{equation}
We see from this decomposition that the right-handed quark $t_R$ can be coupled to the $\mathbf1$ of SO(5)$'$ and the singlet in the $\mathbf5$, while the left-handed doublet $q_L$ can only couple with the $\mathbf5$.
We consider the following Lagrangian for the top quark:
\begin{equation}
\mathcal L_{\text{int}}^f = {\bar q_L}^\alpha\,{\mathcal Y_{L,a}^{\alpha}}^T\,\mathcal O_L^a + \bar t_R\,\mathcal Y_{R,a}^T\,\mathcal O_R^a + \text{h.c.}\,,
\label{eq:Lagrangian_spurion_fermion}
\end{equation}
where $a=1,\dots,7$ is an SO(7) index, $\alpha=1,2$ is the flavor index of the quark doublet and $\mathcal Y_{L,R}$ are the spurions. The SM fermions are assumed to be even under both $C_2$ and $P_7$.

We promote the couplings $\mathcal Y_{L,R}$ to fields (the spurions) whose transformations under $\GG$ are dictated by the ones of the operators $\mathcal{O}_{L,R}$.
Compatibly with $P_7$, and rotating away unphysical components with the elementary $\U(2)_L^{el}$ and $\U(1)_R^{el}$ symmetries under which the spurions and the quark fields transform,
the most general VEVs for the spurions are:
\begin{equation}
\mathcal Y_L=\frac{y_L}{\sqrt2}
{\begin{pmatrix}
0 & 0 & i & 1 & 0 & 0 & 0\\
i & -1 & 0 & 0 & 0 & 0 & 0
\end{pmatrix}}^T
\label{eq:yL_VEV}\,,\qquad \mathcal Y_R = y_R{
\begin{pmatrix}
0 & 0 & 0 & 0 & 0 & \cos\theta_t & i\sin\theta_t
\end{pmatrix}}^T\,,
\end{equation}
with $y_L$ and $y_R$ real, 
and the fifth component of $\mathcal Y_R$ set to zero by the $P_7$ parity. It is evident that the VEV of $\mathcal Y_R$ breaks $C_2$ unless $\theta_t=0$. It is important to distinguish between two types of symmetries: the spurionic ones are symmetries of the strong sector which are unbroken before the spurions acquire a VEV; the residual ones are symmetries at the electroweak scale, which remain unbroken even after the spurions have acquired a VEV. In the following, we assume that spurionic symmetries are respected by the theory.

In order to build the partial compositeness Lagrangian, we ``dress" the spurions with the NGB matrix and define:
\begin{subequations}
\begin{align}
\label{eq:dressed_Y_L}
&\bar{\mathcal Y}_L^\alpha \equiv {\left(r_\theta^{-1}U_\theta^\dagger\,\mathcal Y_L\right)}^\alpha\,,\\
\label{eq:dressed_Y_R}
&\bar{\mathcal Y}_R \equiv r_\theta^{-1}U_\theta^\dagger\,\mathcal Y_R\,.
\end{align}
\end{subequations}
This definition is consistent with the standard one, i.e. $\bar{\mathcal Y}=U^\dagger\mathcal Y$: this can be easily checked by going to the basis of the VEVs, where $\langle U_\theta\rangle=0$ and $\langle U\rangle=r_\theta$.

In general, the dressing procedure has the effect to take an object transforming with an index $a$ of $\mathcal G$ into a new object transforming with an index $\bar a$ of $\mathcal H$. It is then understood that whenever a barred quantity appears, barred indices are implicit.

In order to write the low-energy effective Lagrangian obtained from integrating out the composite sector, we follow the standard procedure, detailed for example in ref.~\cite{Mrazek:2011iu}: we use the dressed spurions and SM fields to write operators invariant under $\mathcal H$. This will also assure their invariance under the full $\mathcal G$. The pNGB dependence will be included in the dressed spurions.

In the aligned limit, $\theta_{1,2}=0$, the dressed spurions transform as a $(\mathbf5,\mathbf1) \oplus (\mathbf1,\mathbf2)$ under SO(5)$\times$SO(2), with components given by:
\begin{equation}
\left(\bar{\mathcal Y}_5\right)^i \;,\; \left(\bar{\mathcal Y}_2\right)^\mu \;,\;
\end{equation}
with $i=1,\dots,5$, while $\mu=1,2$ is the index associated to SO(2). The effective Lagrangian can then be constructed by combining them with $\delta_{ij}$, $\delta_{\mu\nu}$ and $\epsilon_{\mu\nu}$. The latter possibility, however, violates the $C_2$-spurionic and will not be considered. In addition, only left-right combinations have to be considered because of chirality. Finally, the two invariants which can be constructed with $\delta$ symbols are not independent, due to the singlet one can obtain by combining two $\mathbf7$.

The leading order operator generating the top mass is thus given by:
\begin{equation}
\mathcal L_t = c_t \frac{m_*}{g_*^2}\,\bar q_L^{\,\alpha}\left(\bar{\mathcal Y}_{L,2}^\alpha\right)_\mu^{\dagger} \left(\bar{\mathcal Y}_{R,2}\right)^\mu t_R\,,
\label{eq:effective_Lagrangian_top}
\end{equation}
where the coefficient $m_* / g_*^2$ comes from NDA and is such that $c_t$ is a coefficient expected to be $\OO(1)$.
From this effective Lagrangian one obtains both the top mass and the top-NGBs interactions. Expanding around $\beta = 0$, the top quark Yukawa coupling is given by:
\begin{equation}\label{eq:YukTop}
Y_t \approx c_t \frac{y_L y_R}{g_*}(\sqrt{1-\xi} \cos \theta_t + \beta \sin \theta_t)~.
\end{equation}
Note that the factor in parenthesis approaches $\beta$ for $\theta_t \to \pi/2$. This suppression can be compensated by a slightly larger value of $c_t$ or of $y_L y_R / g_*$.  We can consider an analogue Lagrangian for the compositeness of the other quarks, but since $Y_b\ll Y_t$, the other contributions are expected to be subleading. In particular, given that the choice has no major effect on the potential, we take $\theta_b=0$.

A crucial point is that~\cref{eq:effective_Lagrangian_top} leads to an interaction of the type $i h A_0\bar t\gamma^5t$ after the spurions acquire a VEV: this interaction explicitly violates $C_2$ (in particular, the one which is broken is the $C_2$-residual in the language introduced before), implying that in general also the second doublet takes a VEV.

Furthermore, it also turns out that $\eta$ and $\kappa$ have opposite $CP$-parities: we assume that $\eta$ is even and $\kappa$ is odd.

In order to avoid sizable flavor changing neutral currents (FCNCs), the embedding of all the other SM fermions should be fixed carefully. Choosing the fundamental representation for the fermions embedding leads in principle to two independent strong sector invariants; however, the spurionic $C_2$ forbids one of these. Nevertheless, to avoid FCNCs, different families should have the same embedding, including the same choice of $\theta_u = \theta_c = \theta_t$, $\theta_d = \theta_s = \theta_b$.
Even so, it is well known that in Composite Higgs models operators from the strong dynamics at the energy scale $\Lambda$ can induce potentially dangerous flavor violating effects both in the quark and in the lepton sector. However, a more detailed discussion of the flavor phenomenology of this model is beyond the scope of this work. 

\subsection{Pseudo-NGB potential}
\label{sec:NGB_potential}

The pNGB potential is generated from the explicit breaking of the Goldstone symmetry due to the gauging of the EW subgroup of $\mathcal{G}$ and to the mixing between SM fermions and the composite sector.
Using naive dimensional analysis, the radiatively-generated potential can be written schematically as (see e.g. ref.~\cite{Panico:2011pw}):
\begin{equation}
V(\Pi) \sim m_*^2 f^2 \left( \frac{g_*^2}{16\pi^2} \right)^L \left( \frac{g_{\SM}}{g_*} \right)^{\mu_G} \left( \frac{y}{g_*} \right)^{\mu_F} \hat{V}\left( \frac{\Pi}{f} \right)
\label{eq:higgs_pot_general}
\end{equation}
where $\hat V$ is a dimension-less function of the NGBs, $L$ counts the number of loops at which each term is generated, and $\mu_G$ and $\mu_F$ count the required insertions of the gauge and fermionic spurions, respectively.

The construction of the different terms in the potential by building invariants from the spurions follows closely the discussion presented in \cite{Mrazek:2011iu}. In the following we shortly describe only the main parts.


\subsubsection{Gauge contributions}
\label{sec:pot_gauge}

One source of explicit breaking of the global symmetry of the strong sector are interactions between the SM gauge bosons and the pNGBs.
It is convenient to introduce a set of spurions:
\begin{equation}
\mathcal G'\equiv\sum_{A=1}^{21}\mathcal G'_A\,T^A = g'\,T^R_3\,~, \qquad
\mathcal G^\alpha\equiv\sum_{A=1}^{21}\mathcal G_A^\alpha\,T^A = g\, T^\alpha_L~,
\label{eq:spurion_gauge}
\end{equation}
transforming under $g \in G$ as $\mathcal G^X \to g \mathcal G^X g^\dagger$. They can be dressed with NGBs as:
\begin{equation}
\bar{\mathcal G}^X\equiv r_\theta^{-1}U_\theta^\dagger \mathcal G^X\, U_\theta r_\theta\,.
\label{eq:dressed_G'}
\end{equation}
Their components $\bar{\mathcal G}^X_A = \tr\left[\bar{\mathcal G}^X T^A\right]$ transform as the following multiplets of $\SO(5)\times\SO(2)$:
\begin{equation}
\mathbf{21} = (\mathbf{10},\mathbf 1) \oplus (\mathbf 5,\mathbf 2) \oplus (\mathbf 1,\mathbf 1)\,,
\label{eq:decomp_adjoint}
\end{equation}
associated to $\left\{T_L,T_R,T_5\right\}$, $\{\hat T_1,\hat T_2\}$ and $T_2$, respectively.
We can thus organize the components of $\bar{\mathcal G}^X$ as:
\begin{equation}
\left(\bar{\mathcal G}^X_{10}\right)^I \;,\; \left(\bar{\mathcal G}^X_{\hat T}\right)^i_\mu \;,\; \bar{\mathcal G}^X_2\,,
\end{equation}
with $I=1,\dots,10$ being an index in the adjoint, while $i=1,\dots,5$ and $\mu=1,2$ being the indices associated to SO(5) and SO(2), respectively.
The set of independent invariants with two spurion insertions, compatible with $C_2$ and $P_7$, is:
\begin{align}
\mathcal I^{(1)}_{g'} &\equiv -\left(\bar{\mathcal G}'_{10}\right)^I\left(\bar{\mathcal G}'_{10}\right)_I~,
& \mathcal I^{(2)}_{g'} &\equiv -\bar{\mathcal G}'_2 \, \bar{\mathcal G}'_2 ~, \notag\\ 
\mathcal I^{(1)}_{g} &\equiv -\left(\bar{\mathcal G}^\alpha_{10}\right)^I\left(\bar{\mathcal G}^\alpha_{10}\right)_I~,
& \mathcal I^{(2)}_{g} &\equiv -\bar{\mathcal G}^\alpha_2 \, \bar{\mathcal G}^\alpha_2 ~.
\end{align}
The gauge contribution to the NGB potential is then given by:
\begin{equation}
V_{\rm gauge}=\frac{m_*^4}{16\pi^2}\sum_{i}\sum_{\tilde g=g,g'}\frac{1}{g_*^2}\,c_{\tilde g}^{(i)}\,\mathcal I_{\tilde g}^{(i)}\,,
\label{eq:higgs_pot_gauge}
\end{equation}
where $c_{g, g^\prime}^{(i)}$ are $\mathcal{O}(1)$ coefficients.


\subsubsection{Fermionic contribution}
\label{sec:pot_fermion}

The main source of explicit breaking of the Goldstone symmetry is due to the coupling of the composite sector with elementary quarks, and in particular with the top. The relevant Lagrangian was already introduced in~\cref{eq:Lagrangian_spurion_fermion}.

The first step to build the possible invariants which can enter in the potential is to construct combinations of spurions which are invariant under the elementary gauge symmetry of~\cref{eq:Lagrangian_spurion_fermion}:
\begin{subequations}
\begin{alignat}{3}
\label{eq:Delta_L}
\bar\Delta_L^{\bar a\bar b} &\equiv \bar{\mathcal Y}_L^{*\,\alpha,\bar a}\bar{\mathcal Y}_L^{\alpha,\bar b}\,,\\[.15truecm]
\label{eq:Delta_R}
\bar\Delta_R^{\bar a\bar b} &\equiv \bar{\mathcal Y}_R^{*\,\bar a}\bar{\mathcal Y}_R^{\bar b}\,.
\end{alignat}
\end{subequations}
The independent set of invariants which can be obtained at $\mathcal{O}(y^2)$ are:
\begin{equation}
\mathcal I_{(1,0)}^{(1)}=\bar{\Delta}_L^{ii}\,,\quad \mathcal I_{(0,1)}^{(1)}=\bar{\Delta}_R^{ii}\,.
\label{eq:invariants_order_2}
\end{equation}
At $\mathcal{O}(y^4)$, the non-vanishing invariants are: 
\begin{align}
&\mathcal I_{(2,0)}^{(1)} \equiv \bar\Delta_L^{ij}\,\bar\Delta_L^{ji} \,,\quad \mathcal I_{(1,1)}^{(1)} \equiv \bar\Delta_L^{ij}\,\bar\Delta_R^{ji} \,,\quad \mathcal I_{(0,2)}^{(1)} \equiv \bar\Delta_R^{ij}\,\bar\Delta_R^{ji}\,, \nonumber\\[.2truecm]
&\mathcal I_{(2,0)}^{(2)} \equiv \bar\Delta_L^{ii}\,\bar\Delta_L^{jj} \,,\quad \mathcal I_{(1,1)}^{(2)} \equiv \bar\Delta_L^{ii}\,\bar\Delta_R^{jj} \,,\quad \mathcal I_{(0,2)}^{(2)} \equiv \bar\Delta_R^{ij}\,\bar\Delta_R^{ij}\,, \nonumber\\[.2truecm]
\label{eq:I_fermion_1}
&\mathcal I_{(0,2)}^{(3)} \equiv \Im\left[\bar\Delta_R^{\bar ai}\,\bar\Delta_R^{\bar ai}\right]\,,
\end{align}
where the indices have to be interpreted as already indicated. While the operators indicated with $ ^{(1)}$ are generated at one loop, all the other ones are generated at two loops~\cite{Mrazek:2011iu}, and are thus accompanied by a further factor of $g_*^2/(4\pi)^2$.

The general form of the scalar potential was given in~\cref{eq:higgs_pot_general}; for the fermionic case, it can be expressed as:
\begin{equation}
V_{\rm fermion}=N_c\,\frac{m_*^4}{16\pi^2}\sum_{n_L,n_R,i}\frac{1}{g_*^{2(n_L+n_R)}}\,c_{(n_L,n_R)}^{(i)}\,\mathcal I_{(n_L,n_R)}^{(i)}\,,
\label{eq:higgs_pot_fermion}
\end{equation}
where $\mathcal I_{(n_L,n_R)}^{(i)}$ is an invariant formed with $n_{L,R}$ powers of $\bar{\Delta}_{L,R}$, and $c_{(n_L,n_R)}^{(i)}$ are $\mathcal O(1)$ coefficients. Since the fermions in the loop generating the potential are colored, there is a factor $N_c$ accounting for the number of colors; in the following, we take $N_c=3$. Notice that $\mathcal I_{(n_L,n_R)}^{(i)}\propto y_L^{2n_L}y_R^{2n_R}$, which is the reason for the denominator in the previous estimate for the potential. Since we assumed there is no further $CP$ breaking coming from the effective Lagrangian, we set $c^{(3)}_{(0,2)}$ to 0, since the associated invariant contains $CP$ breaking terms.

It turns out that $c^{(1)}_{(1,0)}$, $c^{(1)}_{(0,1)}$, $c^{(1)}_{(2,0)}$, $c^{(1)}_{(1,1)}$, $c^{(1)}_{(0,2)}$, $c^{(2)}_{(1,1)}$ and $c^{(2)}_{(0,2)}$ are the most relevant coefficients for numerical estimates. In our numerical scans, we take for simplicity all the other coefficients (namely  $c^{(2)}_{(2,0)}$ and the ones coming from the gauge invariants) equal to 1, since they do not play a relevant role. 
We generically denote by $c_i$ the $\mathcal O(1)$ coefficients, and define three possible ranges of variation of these coefficients, depending on how close they are to  unity:
\begin{itemize}
    \item \emph{strictly} natural coefficients: $0.2\leq|c_i|\leq 5$;
    \item \emph{loosely} natural coefficients: $0.1\leq|c_i|\leq 10$;
    \item unnatural coefficients: $|c_i|<0.1$ or $|c_i|>10$.
\end{itemize}

\section{NGB dynamics}
\label{sec:dynamics}

In this section, we summarize the main properties of the pNBGs, such as their vacuum structure, their spectrum and interactions.


\subsection{Vacuum structure}
\label{sec:vacuum}

By setting to zero the pNGB fields in the potential $V_{\rm tot} = V_{\rm gauge} + V_{\rm fermion}$, we can find the minimum for the misalignment angles $\theta_1$ and $\theta_2$. In practice, we impose that the minimum is found for the required benchmark values of $\xi$ and $\beta$ by solving for two of the free coefficients. Specifically, we solve for $c_{(1,0)}^{(1)}$ and $c_{(1,1)}^{(1)}$ and check that the solution lies within the desired naturalness range. 
An approximate expression for $\xi$, obtained at leading order in $y_{L,R}/g_* \ll 1$ is:
\begin{equation}
    \xi = \sin^2 \theta_1 + \sin^2 \theta_2 \approx \frac{ 2 N_c y_L^4 c^{(1)}_{(2,0)}+ g_*^2 \left(N_c y_L^2 c^{(1)}_{(1,0)} - 3 g^2 c_{g}^{(1)} - g^{\prime \, 2} c_{g^\prime}^{(1)} \right)}{N_c y_L^4 c^{(1)}_{(2,0)}}~.
\end{equation}
A tuning among the coefficients in the numerator must be imposed in order to reproduce the desired misalignment, the amount of which is of order $\Delta \sim \xi^{-1}$.
As already discussed in ref.~\cite{Mrazek:2011iu}, a hierarchy $\theta_2 \ll \theta_1$, i.e. $\beta \ll 1$, is instead naturally obtained in this model. By minimizing the potential, we get approximately:
\begin{equation}\label{eq:beta_val}
\tan \beta = \frac{\sin \theta_2}{\sin \theta_1} \approx \frac{N_c\, c^{(1)}_{(1,1)} y_L^2 y_R^2 \sin 2 \theta_t }{2 g_*^2 ( g^{\prime 2} c_{g^\prime}^{(1)}+ 2N_c\, y_R^2 c^{(1)}_{(0,1)} \cos 2\theta_t)}~.
\end{equation}
A strong suppression is automatically obtained for $\theta_t \sim \pi/2$. Furthermore, for $g^\prime \ll y_R \cos 2\theta_t$, one can approximate $\tan \beta \sim y_L^2 / g_*^2 \tan 2 \theta_t$ which shows clearly a suppression if $y_L \ll g_*$. Another interesting region we will study is close to $\theta_t \approx \pi/4$. In this case the hypercharge term in the denominator cannot be neglected, but values of $\tan \beta \sim 0.1$ are still naturally obtained.

\subsection{Spectrum}

Due to the smallness of $\beta$, we can perform a power expansion in the expression for the pNGB masses we obtain from the potential.
A mixing between $h$ and $H_0$ is present in general, and can be diagonalized via a rotation by an angle $\alpha \approx \beta$.
Once the conditions fixing $\xi$ and $\beta$ have been imposed, the physical Higgs mass at leading order in $\xi$ and $\beta$ is given approximately by:
\begin{align}
    m_h^2 &\approx \frac{N_c f^2 \xi}{16\pi^2} \left( 2 y_L^4 c^{(1)}_{(2,0)} + y_R^4 c^{(1)}_{(0,2)} (3 + 4 \cos 2\theta_t + \cos 4 \theta_t) \right)  \notag\\
    &\approx \frac{N_c g_*^2 }{8 \pi^2} m_t^2 \left( 2 \frac{y_L^2}{y_R^2} c^{(1)}_{(2,0)} + \frac{y_R^2}{y_L^2} c^{(1)}_{(0,2)} (3 + 4 \cos 2\theta_t + \cos 4 \theta_t) \right)~,
\end{align}
where we omitted contributions from gauge or two-loop coefficients.
In the second line we substituted the expression for the top-Yukawa (cf.~\cref{eq:YukTop}).
A small value of $g_*$, i.e. light top partners, helps to avoid a further tuning in order to obtain the correct Higgs mass. For this reason in the numerical analysis we fix $g_* = 3$. In practice, we impose the measured value $m_h \approx \SI{125}{\giga\electronvolt}$, by solving the (exact) $m_h$ expression for the coefficient $c^{(1)}_{(2,0)}$. 

\begin{figure}
\centering
\includegraphics[scale=0.45]{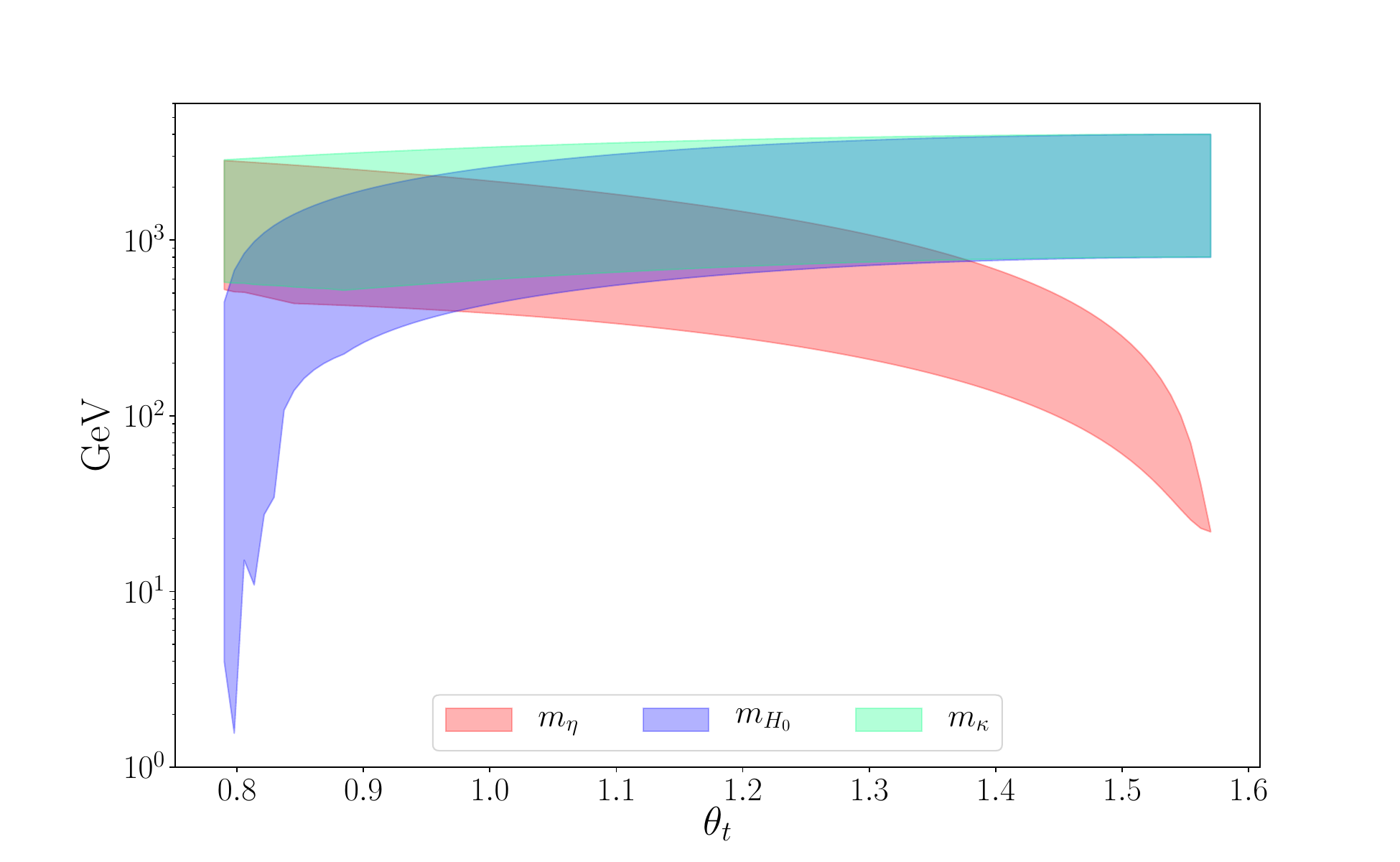}
\caption{\label{fig:mass_spectrum}Mass spectrum for $\xi=0.061$, $\beta=0.1$ and $g_*=3$. The bands are obtained by varying $c^{(1)}_{(0,1)}$, $c^{(1)}_{(0,2)}$, $c^{(2)}_{(0,2)}$ and $c^{(2)}_{(1,1)}$ in the strictly natural range, $|c_i| \in [0.2,5]$.}
\end{figure}

Once $\xi$, $\beta$, and $m_h$ have been fixed, the masses of the other pNGBs, $H_0$, $A_0$, $H_\pm$, $\eta$ and $\kappa$ as functions of the remaining coefficients, to the leading order in $\xi$, are:

\begin{subequations}
\begin{align}
m_{H_0}^2 &\approx  -N_c\frac{y_R^2}{8\pi^2}\left(c_{(0,1)}^{(1)}+\frac{y_L^2}{8\pi^2} c_{(1,1)}^{(2)}\right)m_*^2\cos(2\theta_t)\,, \\[.15truecm]
m_{A_0}^2 &\approx  m_{H_0}^2\,, \\[.15truecm]
m_{H_\pm}^2 &\approx  m_{H_0}^2 - \frac{m_h^2}{2}\,,  \\[.15truecm]
\label{eq:m_eta}
m_\eta^2 &\approx  N_c\frac{y_R^2}{8\pi^2}\left(c_{(0,1)}^{(1)}+\frac{y_L^2}{8\pi^2} c_{(1,1)}^{(2)}\right)m_*^2\cos^2\theta_t \,, \\[.15truecm]
m_\kappa^2 &\approx N_c\frac{y_R^2}{8\pi^2}\left(c_{(0,1)}^{(1)}+\frac{y_L^2}{8\pi^2} c_{(1,1)}^{(2)}\right)m_*^2\sin^2\theta_t\,.
\end{align}
\end{subequations}
Due to our choice of coefficients the gauge contribution cancels at the first order in $\xi$, but it is present at the next to leading one.
As we can see, $H_0$, $A_0$ and $H_\pm$ are almost degenerate in mass. We then assume that $\pi/4 \leq \theta_t \leq \pi/2$ and $c_{(0,1)}^{(1)},c_{(1,1)}^{(2)}>0$; with this choice, all the mass parameters are positive. The sperum is shown in~\cref{fig:mass_spectrum} for $\xi=0.061$ and $\beta=0.1$, as well as different values of $c^{(1)}_{(0,1)}$, $c^{(1)}_{(0,2)}$, $c^{(2)}_{(0,2)}$ and $c^{(5)}_{(1,1)}$.

When $\theta_t\approx\pi/2$, also the first order in $\xi$ can play an important role for $m_\eta$; if all the coefficients are $\mathcal O(1)$, we can approximate it as:
\begin{equation}
m_\eta^2 \approx N_c\, c_{(0,1)}^{(1)}\frac{y_R^2 }{8\pi^2}m_*^2\cos^2\theta_t + \xi\,\frac{m_h^2}{2}\,.
\end{equation}
From~\cref{fig:mass_spectrum}, we notice two main interesting regions to study in more detail:
the first one is for $\theta_t \approx \pi/2$; in this case, $\eta$ is by far the lightest pNGB (other than the SM Higgs), with a mass $m_\eta \sim \mathcal{O}(\SI{100}{\giga\electronvolt})$. All other pNGBs have $\mathcal O(\SI{1}{\tera\electronvolt})$ masses and do not participate in a relevant way to the phenomenology.
The second region is for $\theta_t \gtrsim \pi/4$ and with $\eta$ and $\kappa$ very close in mass, of $\mathcal{O}(\SI{1}{\tera\electronvolt})$; in this case, if $\kappa$ has a long enough lifetime, it can freeze-out in the early universe and then decays into $\eta$, which is the stable DM relic, giving rise to non-thermal DM production.

These two scenarios are studied in~\cref{sec:ThDM,sec:non_thermal_DM}, respectively.

\subsection{pNGB interactions}
In order to study the pNGBs interactions, it is convenient to consider separately those coming from CCWZ and the ones coming from partial compositeness and the potential.

\subsubsection*{Interactions from CCWZ}

The CCWZ Lagrangian in~\cref{eq:LCCWZ} contains pNGB interactions with the SM EW gauge fields, as well as derivative self-interactions:
\begin{align}
\mathcal L_{\Pi}^{(2)} \supset& \mathcal{L}_{\rm kin} + \left(m_W^2W_\mu^+W^\mu_-+\frac{m_Z^2}{2}Z_\mu Z^\mu\right)\left( 1 + \frac{2 g_V}{v} h + \frac{b_h}{v^2} h^2 +\frac{\lambda_\eta^{(V)}}{2v^2} \eta^2 +\frac{\lambda_\kappa^{(V)}}{2v^2} \kappa^2 + \ldots \right) \nonumber\\ 
& - \frac{2}{v}\left( g_{H_0W}m_W^2W_\mu^+W^\mu_- + \frac{g_{H_0Z}m_Z^2}{2}Z_\mu Z^\mu\right)H_0 \nonumber\\
& - \frac{k_\mathrm{der}}{4v^2}\bigg[ \left( \Phi_1^2 (\partial_\mu \Phi_1)^2 - (\Phi_1 \partial_\mu \Phi_1)^2\right) +
\left( \Phi_2^2 (\partial_\mu \Phi_2)^2 - (\Phi_2 \partial_\mu \Phi_2)^2\right) + \nonumber\\
& \quad\qquad 2 \left( \Phi_1 \Phi_2 (\partial_\mu \Phi_1 \partial^\mu \Phi_2) - (\Phi_1 \partial_\mu \Phi_2)(\Phi_2 \partial_\mu \Phi_1) \right)  + \left(\Phi_1 \stackrel{\leftrightarrow}{\partial}_\mu \Phi_2  \right)^2 \bigg] \nonumber\\\label{eq:CCWZLagrExpanded}
& - \frac{m_W m_Z}{2v} g_{H_+V} W_\mu^- Z^\mu H_+ + \mathrm{h.c.}  + \ldots
\end{align}
where $\mathcal{L}_{\rm kin}$ contains the pNGB kinetic terms, $k_{\rm der} = 2 \xi /3$, and $\Phi_{1,2}$ are the two pNGB five-plets of \emph{physical} fields introduced in~\cref{eq:PNGB5plets}. We neglected all couplings which break custodial symmetry and which become negligible once the limits from EWPTs are taken into account, and omitted other interactions with two gauge bosons and two pNGBs, less relevant for the phenomenology discussed in the following.

It is worth noticing that interactions with three pNGBs and two derivatives, such as $\eta^2 h$, are absent from the Lagrangian above. This might seem surprising at first, since such interactions have been long known to be present in similar scenarios, and their relevance has been often stressed (see e.g. refs.\cite{Frigerio:2012uc,Marzocca:2014msa,Fonseca:2015gva,Balkin:2018tma}). Such interactions usually arise from $\Pi^4$ terms, once the Higgs(es) takes a VEV. However, since we employ the description of pNGB fields from the misaligned vacuum (as discussed in~\cref{sec:coset}) no field takes a VEV and therefore these terms are not generated.
Another way to easily understand their absence is to set to zero the gauge and Yukawa couplings. In this case, the global symmetry is exact and all vacua are degenerate, so that the $\SO(5)$ and $\SO(5)^\prime$ subgroups are physically equivalent. In this limit, the two-derivative NGBs interactions start at $\mathcal{O}(\Pi^4)$ in both vacua (and are of the form specified above). Now, switching on the gauge and Yukawa couplings selects $\SO(5)^\prime$ as the true vacuum; however, since derivative interactions do not depend on these couplings and since in our descriptions fields do not take a VEV, the derivative interactions are not affected and therefore cubic ones are not generated.

The connection with the description most commonly employed in the literature (i.e. describing fields from the \emph{gauge} vacuum $\SO(5)$ and allowing then the Higgs to take a VEV) can be easily obtained via a non-linear field redefinition (\cite{Mrazek:2011iu,Fonseca:2015gva}). For example, in the limit of $\theta_2 = 0$ this is given by ($\tilde h$ and $\tilde \eta$ are the physical fields in the \emph{gauge} description):
\begin{eqnarray}
h &\to& \tilde h + \theta_1 \dfrac{\tilde \eta^2}{3f} + \mathcal{O}(\theta_1^2) 
\,,\\
\eta &\to& \tilde \eta - \theta_1 \dfrac{\tilde\eta \tilde h}{3f} + \mathcal{O}(\theta_1^2)\,.
\end{eqnarray}
Such a transformation generates cubic derivative interactions from the kinetic terms, as well as non-derivative interactions from the pNGBs mass terms. The net effect of these is to keep physical observables invariant under such transformations.


\subsubsection*{Interactions with fermions and from the potential}

Let us now list other phenomenologically relevant pNGB interactions, in particular those with SM fermions and self-interactions from the potential:
\begin{subequations}
\begin{align}
\mathcal L_q &\supset - \frac{m_q}{v}\,\bar q q \left( k_q h + k_{H_0 q} H_0 - \frac{g_q}{2v^2} \eta^2 + \frac{g_{\kappa q}}{2v^2} \kappa^2 \right)   -\frac{g_{\eta\kappa q}}{v^2}\,m_q \eta\,\kappa\,\bar q\gamma^5q \label{eq:NGBfermCoupl}\\[.2truecm]
 \mathcal{L}_{g}^{t-\text{loop}} &\supset \frac{g_{gh}}{v}hG_{\mu\nu}^aG_a^{\mu\nu}+\frac{g_{gH_0}}{v}H_0G_{\mu\nu}^aG_a^{\mu\nu}+\frac{g_{g\eta}}{v^2}\eta^2G_{\mu\nu}^aG_a^{\mu\nu}\label{eq:NGBgluonCoupl}\\[.2truecm]
V &\supset  - \frac{g_{\eta h}}{2}\,v\,\eta^2h- \frac{g_{\eta H_0}}{2}\,v\,\eta^2H_0  - \frac{g_{\kappa h}}{2}\,v\,\kappa^2h - \frac{g_{\kappa H_0}}{2}\,v\,\kappa^2H_0 \notag\\[.2truecm]
&\quad - \frac{\lambda_{\eta h}}{4}\,\eta^2h^2 - \frac{\lambda_{\eta H_0}}{4}\,\eta^2H_0^2 + \frac{\lambda_{\eta A_0}}{4}\,\eta^2A_0^2 + \frac{\lambda_{\eta H_+}}{2}\,\eta^2H_+H_- \notag\\[.2truecm]
&\quad - \frac{g_{A_0h}}{2}\,v\,h\,A_0^2 - \frac{g_{A_0H_0}}{2}\,v\,H_0A_0^2 - g_{H_+h}v h\,H_+H_- - g_{H_+H_0}v H_0H_+H_- \notag\\[.2truecm]
&\quad + \frac{m_h^2}{2v^2}\,\lambda v\,h^3 - \frac{\lambda_{H_0}}{6}\,v\,H_0^3 + \frac{g_{H_0}}{2}\,v\,h\,H_0^2 - \frac{g_{H_0hh}}{2}\,v\,h^2H_0 \notag\\[.2truecm]
&\quad - \frac{\lambda_{\kappa h}}{4}\,\kappa^2h^2 - \frac{\lambda_{\kappa H_0}}{4}\,\kappa^2H_0^2 + \frac{\lambda_{\kappa A_0}}{4}\,\kappa^2A_0^2 + \frac{\lambda_{\kappa H_+}}{2}\,\kappa^2H_+H_-
\label{eq:NGBNGBCoupl}
\end{align}
\end{subequations}
The expressions of the effective couplings are reported in~\cref{app:effective_coupl}. We require them to be always less than $4\pi$ for perturbative reasons: this usually forces $c^{(1)}_{(0,1)}$ to be smaller than 1.

\section{Thermal dark matter scenario}
\label{sec:ThDM}

The first region of interest is the one for $\theta_t\lesssim\pi/2$, where $\eta$ is the lightest pNGB and the dark matter abundance is generated via a thermal freeze-out. The fact that $\theta_t\approx\pi/2$ has several implications:
from~\cref{fig:mass_spectrum}, we see that $\eta$ is much lighter than the other resonances, so that the effect of other pNGBs can be largely neglected for the freeze-out computation; a small value of $\beta$ is natural and does not require a further tuning or unnaturally small value of the potential coefficients, see~\cref{eq:beta_val}; finally, since $\cos\theta_t$ is small, the expression for the top mass,~\cref{eq:YukTop}, has a mild suppression which must be compensated by having either a large $c_t$ or $y_Ly_R\gtrsim g_*$; we choose the second option for naturalness reasons, and take $y_L=2$, $y_R=3$ and $g_*=3$. This choice is consistent with perturbativity and with the phenomenological requirement of having a less composite left-handed top rather than the right-handed one. Other choices are possible and we do not expect them to change the results qualitatively.

We also explored other regions of the parameter space: in particular, the region in the range $\theta_t\approx\pi/4$ could be potentially interesting because of coannihilations  with other pNGBs (cf.~\cref{fig:mass_spectrum}); however, we verified that in this region it is not possible to reproduce the observed relic density via the standard freeze-out mechanism.
Instead, in this region a non-thermal DM production mechanism can take place, whose discussion is deferred to~\cref{sec:non_thermal_DM}.

\subsection{Relic density}

\begin{figure}
\begin{center}
\includegraphics[scale=0.5]{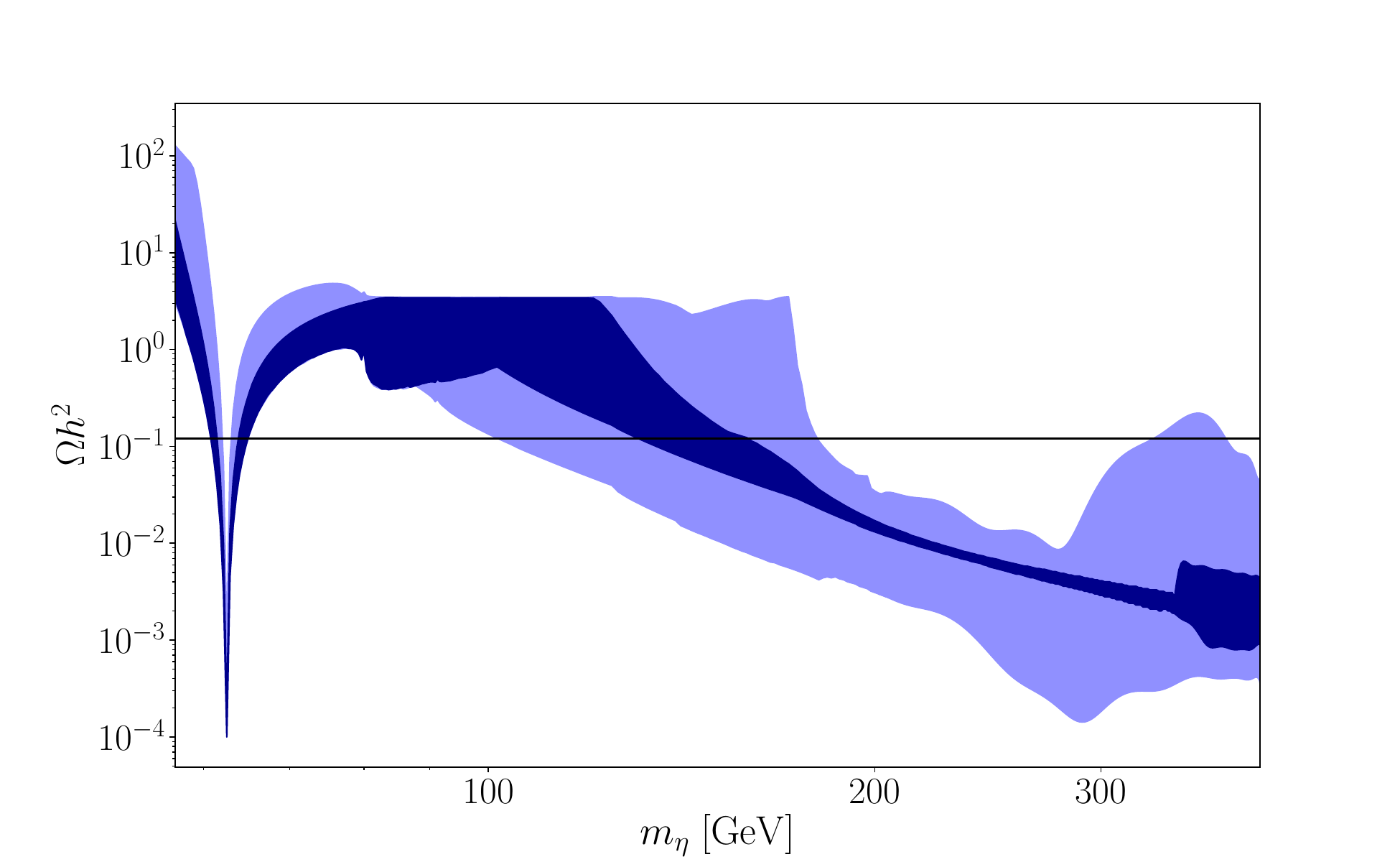}
\end{center}
\caption{\label{fig:relic_density_th} Relic density as a function of the 
DM mass $m_\eta$, for $f=1 \, \TeV$ ($\xi=0.061$) and $\beta=0.1$; the black line corresponds to the measured value $\Omega h^2=0.1198$~\cite{Aghanim:2018eyx}. The dark (light) blue region is obtained by letting the coefficients in the potential vary within the strictly (loosely) natural range (see~\cref{sec:pot_fermion}).}
\end{figure}

The main contributions to the relic abundance are given by DM annihilations into SM EW gauge bosons, Higgs and top quark. In our computations, we also included subleading contributions, and all the details can be found in~\cref{app:relicdensity}. The relic density profile as a function of the DM mass is shown in~\cref{fig:relic_density_th}. The darker (lighter) region is obtained by letting the coefficients vary inside the strictly (loosely) natural range.

It is useful to remember that the relic density is inversely proportional to the integrated thermally-averaged annihilation cross section: $\Omega h^2\appropto 1/\langle\sigma v\rangle$. This is the reason why a plateau appears for $m_\eta\lesssim m_h$, where the cross section is dominated by annihilations into SM gauge bosons: since the latter do not depend on the $c_i$'s, the annihilation cross section is always bounded from below. The situation changes at larger masses, where new annihilation channels open up: in principle, we expect the relic density to decrease with increasing mass. However, it is possible that a cancellation in the main contributions to the effective cross section occurs: this is precisely the case in the region $m_\eta\approx\SI{400}{\giga\electronvolt}$ of~\cref{fig:relic_density_th}, where $\eta$ is sufficiently heavy so that the exchange of $H_0$ compensates the exchange of $h$ in $s$-channel, the two contributions having opposite signs.

Of all the effective couplings, the one that plays the most important role is $g_{\eta h}$, describing the interactions between two $h$'s and one $\eta$. It enters different processes with different signs, so it is non-trivial to describe its role analytically. The results highly depend on the parameters $\xi$ and $\beta$ and, guided by EWPTs, we decide to focus our attention on $\xi=0.061$ (corresponding to $f=\SI{1}{\tera\electronvolt}$) and $\beta=0.1$.
From~\cref{fig:relic_density_th}, we see that there are two good mass regimes which give the correct relic density, $\Omega h^2=0.1198$~\cite{Aghanim:2018eyx}: $m_\eta\approx m_h/2$ and $m_\eta\approx\SI{150}{\giga\electronvolt}$; a third one, at $m_\eta\approx\SI{400}{\giga\electronvolt}$, also reproduces the correct relic density if the coefficients are allowed to vary inside the loosely natural range. However, as we will show, this high mass range is already excluded by direct detection.

\subsection{LHC searches}

In the region of parameter space where $m_\eta\leq\SI{62.5}{\giga\electronvolt}$, the Higgs can decay into two DM particles, $h\to\eta\eta$. Experimental constraints on invisible Higgs decays have been obtained by ATLAS and CMS \cite{Sirunyan:2018owy, Aaboud:2019rtt} for various possible assumptions on the Higgs couplings to SM particles.
Since the couplings of the Higgs to quarks and gauge bosons in our model are different from the SM ones, the widths of the decays into SM particles have to be appropriately rescaled. However, no major departure from the SM result is expected for the values of $\xi$ and $\beta$ allowed by the electroweak precision test. We thus take as an experimental bound $BR_\text{inv}<19\%$ at 95\% CL
\cite{Sirunyan:2018owy}. 
The HL-LHC will reach a 95\% CL exclusion sensitivity of $1.9\%$, while future electron-positron colliders would be able to reach sub-percent precision (see, e.g., ref.~\cite{deBlas:2019rxi} for a recent review of Higgs boson measurements at future colliders).

\begin{figure}
\centering
\includegraphics[width=0.7\textwidth]{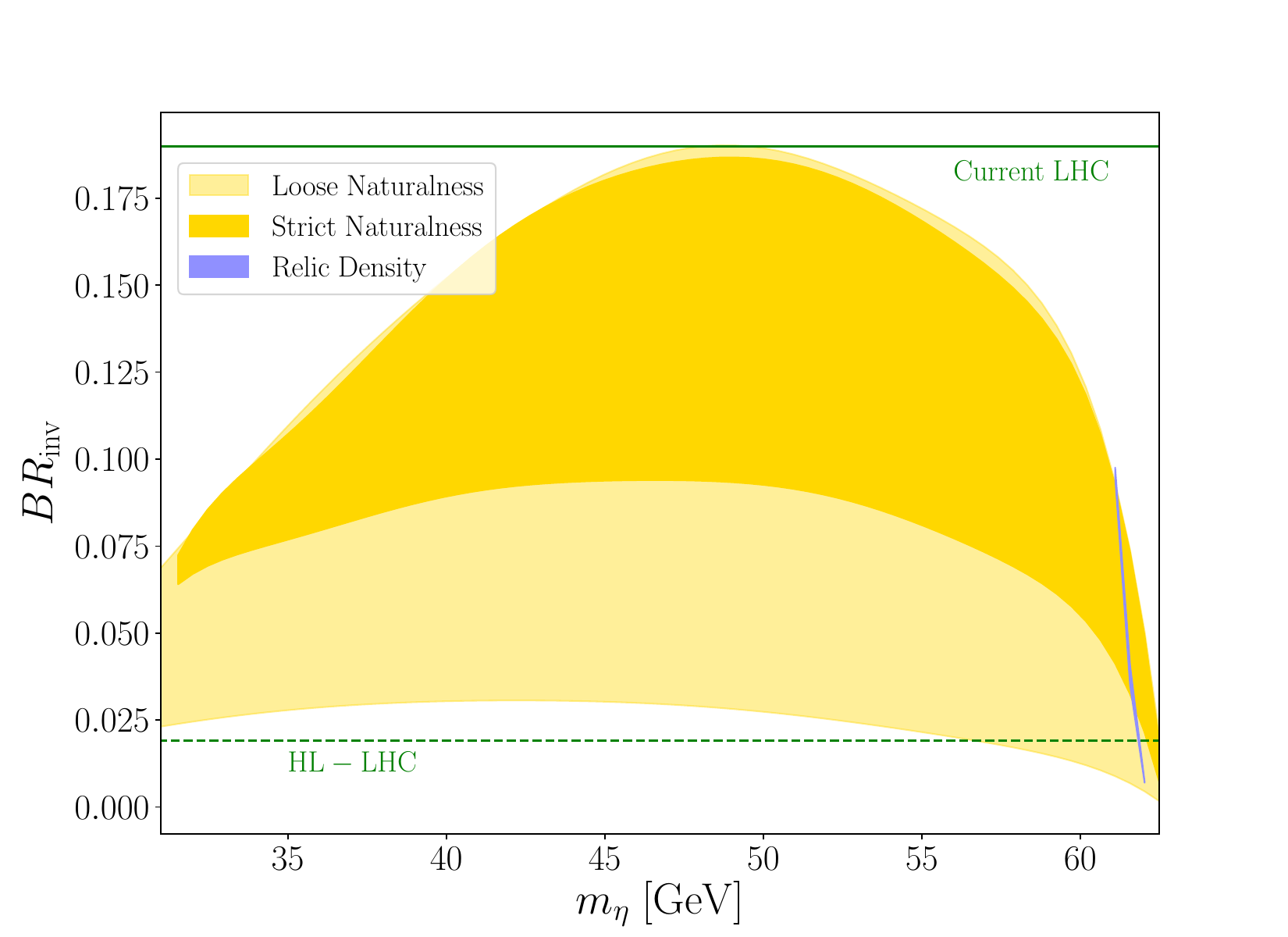}
\caption{\label{fig:Higgs_width}Branching ratio of invisible Higgs decays as a function of the DM mass. The dark (light) yellow region corresponds to strictly (loosely) natural $\mathcal O(1)$ coefficients. The current bound, $BR_\mathrm{inv}<0.19$
at 95\% CL \cite{Sirunyan:2018owy}, is shown as a green solid line, while the HL-LHC prospect is shown with the dashed line. Finally, the region of the parameter space where the observed relic density is reproduced is shown in blue.}
\end{figure}

In our model, the invisible Higgs decay width is given by:
\begin{equation}
\label{eq:Higgs_width}
\Gamma_{h\to\eta\eta} = \frac{g_{\eta h}^2}{32\pi m_h} v^2\sqrt{1-\frac{4m_\eta^2}{m_h^2}}\,,
\end{equation}
which depends on the coefficients $c_i$'s via the effective coupling $g_{\eta h}$.
The corresponding prediction for the invisible branching ratio of the Higgs, obtained by letting the $c_i$'s vary inside the strictly (loosely) natural range are shown as dark (light) yellow regions in~\cref{fig:Higgs_width}. As can be seen, the model's predictions are always lower than the current experimental bound. In the same figure, the small region where also the correct relic density is obtained is shown in blue. It is interesting to notice that HL-LHC will be able to test most of this parameter space via invisible Higgs decays.

The missing energy trace could also be produced by direct production of $\eta$ particles or by the decay of other massive scalars (for instance $H_0$, which is also linearly coupled to $\eta$). In this case, we need to look at specific tags: in our model, the most relevant ones will be an energetic jet, monojet signature (MJ), or two well-separated jets, Vector Boson Fusion (VBF) signature. In both cases, it is important to describe the effective coupling of gluons to the massive scalars,~\cref{eq:NGBgluonCoupl}, since this gives the main contribution to the production cross section.

We implemented the model in  \textsc{FeynRules} \cite{Alloul:2013bka,Christensen:2008py} and generated simulated events with \textsc{MadGraph5} \cite{Alwall:2014hca}.
Even tough $h$ is always lighter than $H_0$ in the parameter space we are considering, diagrams involving $H_0$ still need to be considered in the MJ and VBF processes, since these can give sizable contributions. 
We obtain that for masses above $\SI{50}{\giga\electronvolt}$ both monojet and VBF do not put any constraint on the model, being always at least an order of magnitude below the experimental limits (\cite{Aaboud:2017phn,Aaboud:2018sfi}), for any reasonable values of $\xi$, $\beta$ and the $c_i$.

Overall, DM searches at LHC do not put important constraints on the parameter space of our model.

\subsection{Direct detection}

Direct detection experiments usually put strong constraints on DM models, including the one studied here. In our model, the DM candidate $\eta$ can interact with quarks either via the contact interaction generated by partial compositeness, proportional to the coupling $g_q$  (see~\cref{eq:NGBfermCoupl}), or through an exchange of $h$ and $H_0$. A convenient way to evaluate the spin-independent DM-nucleon cross section is to parametrize the interaction Lagrangian as:
\begin{equation}
\mathcal{L}_{DD}^{(\mathrm{eff})}=\sum_q a_q m_q\eta^2 \bar{q}q\,,
\end{equation}
where:
\begin{equation}\label{eq:DD_amplitude}
a_q=\frac{1}{2}\left[\frac{g_q}{v^2}-\left(k_q\frac{g_{\eta h}}{m_h^2}-k_{H_0q}\frac{g_{\eta H_0}}{m_{H_0}^2}\right)\right].
\end{equation}
As already stressed, in order to avoid FCNCs, one must have $a_t=a_c=a_u$ and $a_b=a_s=a_d$. Because of the different signs between the first term and the parenthesis, it is their interplay to determine the allowed parameter space. In particular, since $g_q$ is negative (see~\cref{eq:eff_coefficients_NGB_fermions}), negative values of the parenthesis will be favored, leading to a partial cancellation of the two terms in~\cref{eq:DD_amplitude}. 
Notice that the value of $a_q$ actually depends on the coefficients $c_i$, so that a cancellation in the scattering amplitude can occur, allowing to evade the DD constraint. From the effective Lagrangian above we derive the spin-independent DM-nucleon cross section \cite{DelNobile:2013sia}. At present the strongest constrain on it comes from the XENON1T experiment \cite{Aprile:2017iyp}.
As can be seen in~\cref{fig:relic_thermal}, this casts important constraints on all the three regions in $m_\eta$ where the relic density is realized in our model, in particular excluding completely the large-mass one.

In our plots we also show the prospects for the future exclusion bounds coming from the XENONnT experiment \cite{Aprile:2015uzo}. We observe that the majority of the parameter space, for the values of $\xi$ and $\beta$ used, will be tested.

\subsection{Indirect detection}
While being somewhat beyond the scope of the paper, we briefly consider the constraints from indirect detection as well. We mainly focus on limits from dwarf spheroidal galaxies (dSphs) given by Fermi-LAT, and reported in ref.~\cite{Ackermann:2015zua}; the relevant one for our model is given by DM-annihilation into $b\bar b$. In the region $m_\eta\approx m_h/2$, only the direct process $\eta\eta\to b\bar b$ has to be considered; for higher DM masses, instead, also the intermediate productions of $W,\,Z,\,h$ (and possibly other NGBs) are important: as a conservative estimate, we assume that these intermediate states completely decay into $b\bar b$. We also took into account possible branching ratios, but the result is only slightly modified by this correction.

As already stated in ref.~\cite{Balkin:2018tma}, also 
anti-protons bounds from AMS-02 are worth exploring;
however, since the size of the systematic uncertainties is
still unclear,  we limit ourselves with  constraints from dSphs, leaving 
a comprehensive treatment of ID in this model to future work.


\subsection{Discussion}
\begin{figure}[t!]
\centering
\includegraphics[width=0.7\linewidth]{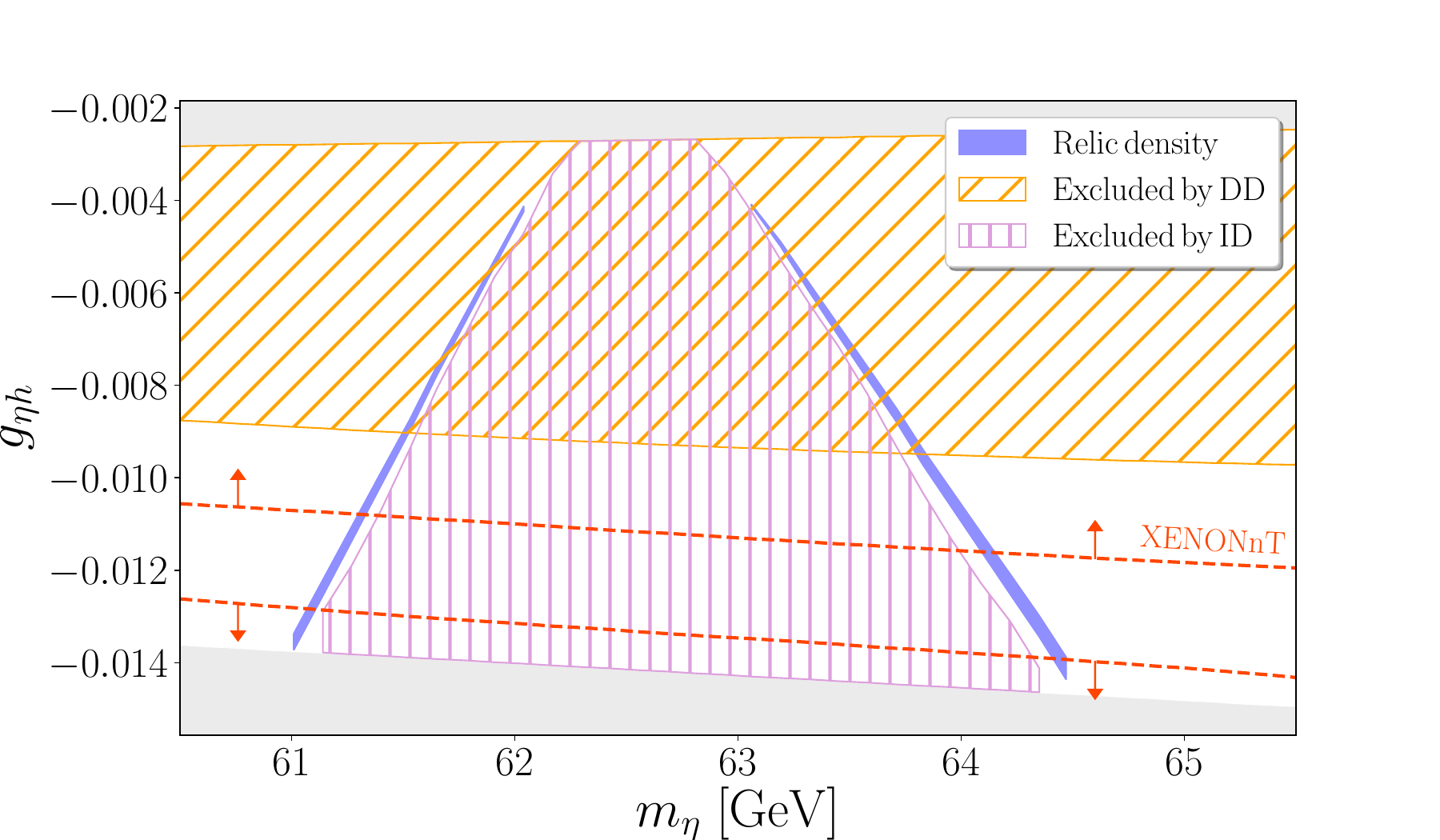}
\includegraphics[width=0.7\linewidth]{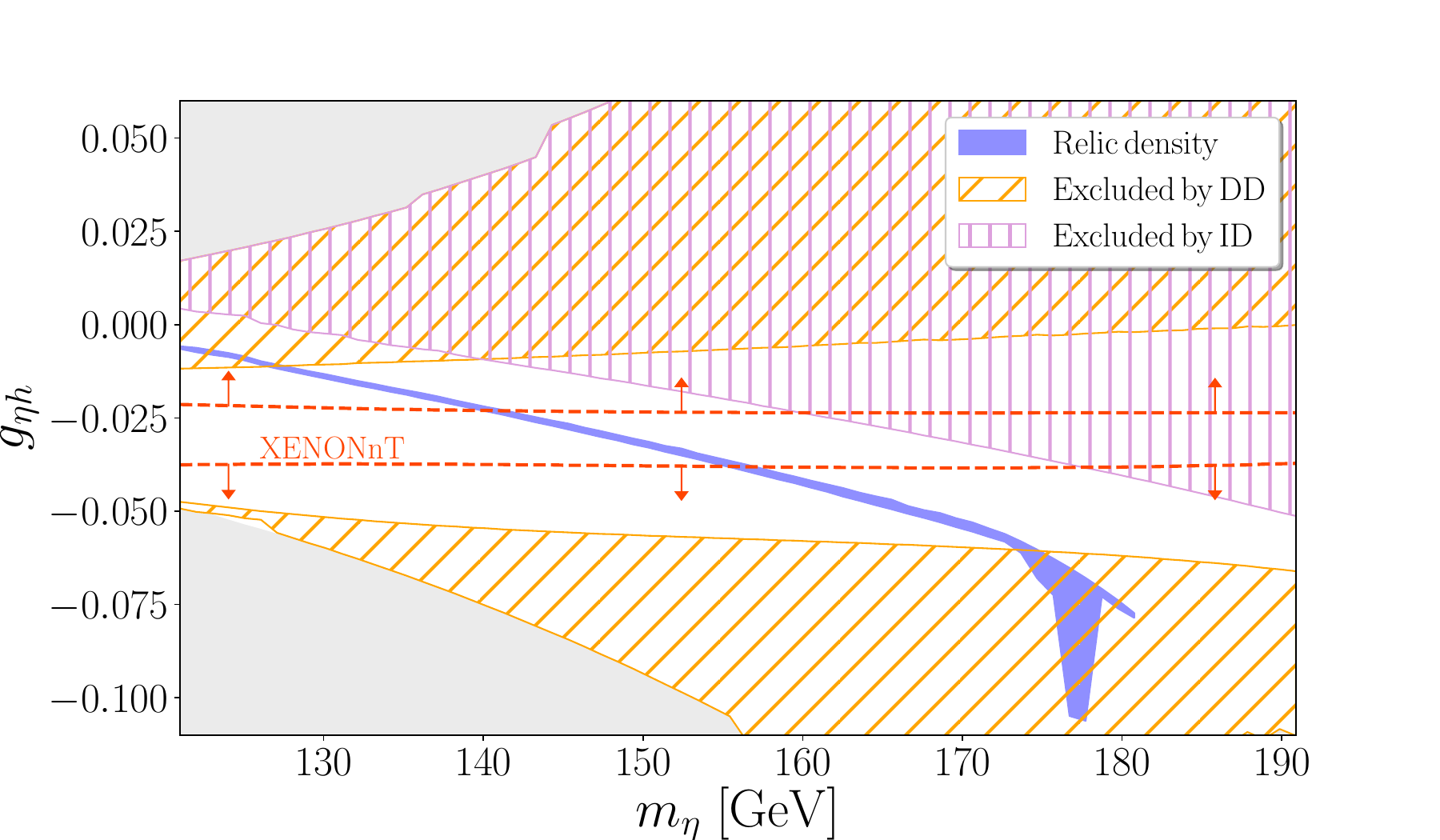}
\includegraphics[width=0.7\linewidth]{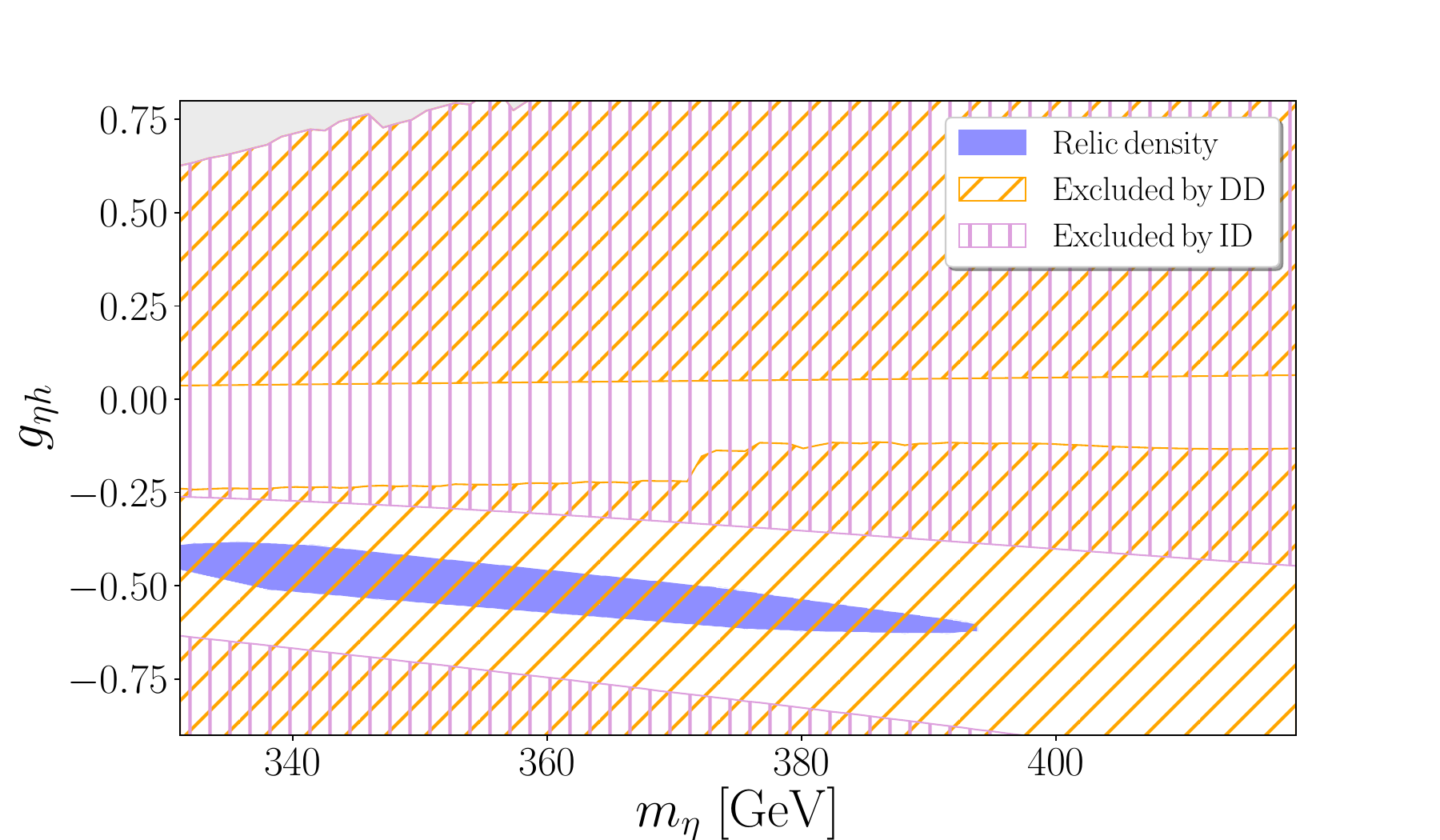}
\caption{\label{fig:relic_thermal} Combination of different features in the $m_\eta$-$g_{\eta h}$ plane for $\xi=0.061$, $\beta=0.1$ and loosely natural $\mathcal O(1)$ coefficients. The blue region corresponds to the $3\sigma$-relic density contour; the orange and purple hatched regions are excluded by direct and indirect detection, respectively; finally, in the gray region the $\mathcal O(1)$ coefficients are outside the range $[0.1,10]$.}
\end{figure}

We summarize the main results for $f=1\,\TeV$ ($\xi=0.061$) and $\beta=0.1$ in~\cref{fig:relic_thermal}.
As anticipated, three regions are possibly interesting: $m_\eta\approx m_h/2$, $m_\eta\approx\SI{150}{\giga\electronvolt}$ and $m_\eta\approx\SI{400}{\giga\electronvolt}$. The orange (purple) hatched area is excluded by DD (ID), while in the blue one the correct DM abundance at $3\sigma$, $\Omega h^2 = 0.1198\pm0.0036$~\cite{Aghanim:2018eyx}, is reproduced.  In these plots, for better readability, we only show the regions obtained by letting the coefficients vary within the loosely natural range. In the gray region the coefficients are beyond this range (unnatural).

Current indirect detection limits from dSphs are close to parameter space where the correct relic density is reproduced, but unable to exclude any portions of it. A future stronger bound should be able to test this model.

\subsubsection*{Low mass range ($\boldsymbol{m_\eta\approx m_h/2}$)}

The correct relic density is reproduced for masses just below and just above the on-shell Higgs production threshold of $\SI{62.5}{\giga\electronvolt}$. Since this regime is good due to the Higgs resonance, the allowed mass range is very narrow and given that there is no symmetry or dynamical argument to expect such a value for $m_\eta$, this would represent a further tuning in the model parameters.
The experimental results coming from the Higgs invisible BR and ID do not exclude any parts of the region where the correct relic density is obtained, while DD sets an upper bound on $g_{\eta h}$.  Even if ID does not exclude any points in the parameter space at the current state, the cross section in our model is only an $\mathcal O(1)$ factor below the experimental constraint, so that upgraded searches are expected to either find a positive result or exclude this region of the parameter space. It is also worth mentioning that HL-LHC will test most of the region below the Higgs pole via a precise measurement of the invisible branching ratio of the Higgs; however, in the near future DD and ID seem more promising directions.

\subsubsection*{Intermediate mass range ($\boldsymbol{m_\eta\approx 150\:\mathrm{GeV}}$)}

At masses larger than around $\SI{100}{\giga\electronvolt}$, the most relevant constraints come from direct detection and the requirements of naturalness of the coefficients. Limits from LHC experiments are not relevant for this mass regime and are not expected to be able to probe it in the near future. EWPTs and Higgs couplings constraints are safe because of the values of $\xi$ and $\beta$ we considered. We are intentionally overestimating the bound from ID, and yet this search is not putting constraints on the model. Upper and lower bounds on $m_\eta$ are set by DD results (cf.~\cref{fig:relic_thermal}). If we were to limit to strictly natural coefficients, then masses below $\SI{135}{\giga\electronvolt}$ would be excluded. The feature at $m_\eta\approx\SI{180}{\giga\electronvolt}$ is because of a cancellation in the cross section for $\eta\eta\to t\bar t$, due to the different sign between the effective couplings $g_{\eta h}$ and $g_{\eta H_0}$ (cf.~\cref{eq:ann_into_top}).

This seems to be the most promising mass range, because ID is pretty weak and DD leaves a significant region of the parameter space available. Also, such $m_\eta$ values are naturally obtained for $\theta_t \lesssim \pi/2$.

\begin{figure}[t]
\centering
\includegraphics[scale=0.4]{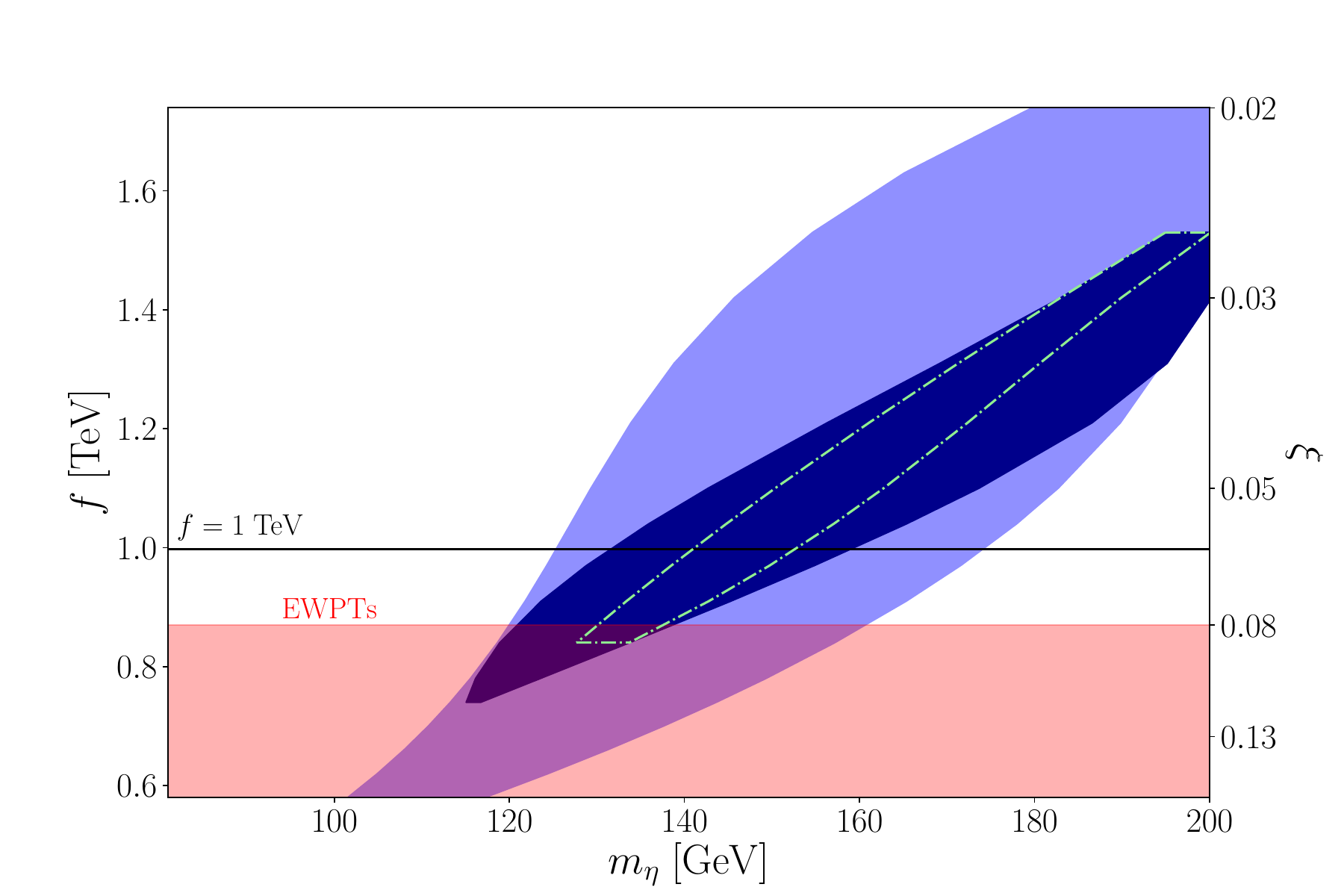}
\caption{\label{fig:relic_mass_f}
The dark (light) blue region represents values of $m_\eta$ and $f$ for which it is possible to reproduce the correct relic density at $3\sigma$, evade DD constraints, all while having strictly (loosely) natural coefficients $c_i$.
The red region is excluded by our combined fit of EWPTs and Higgs couplings (see~\cref{app:EWPTHiggs}), while the black horizontal line corresponds to the benchmark value of $f=1$ TeV  ($\xi = 0.061$) we consider. The green, dashed region correspond to the projection with DD limits from XENONnT and strictly natural coefficients.}
\end{figure}

Decreasing the value of $\xi$ has the effect of enlarging the allowed mass range, so that it is interesting to investigate how the latter varies with varying fine tuning: this can be seen in~\cref{fig:relic_mass_f}.
We observe that DM phenomenology in this model allows for low values of $f$, up to $f \approx \SI{750}{\giga\electronvolt}$ for strictly natural coefficients and even below $\SI{600}{\giga\electronvolt}$ for loosely natural ones. Indeed, the relevant lower bound on $f$ in our setup is the one due to EWPTs and Higgs coupling measurements.
This should be compared to other similar non-minimal composite DM models in the literature, for which significantly larger values of $f$ were found to be necessary (see e.g. \cite{Fonseca:2015gva, Balkin:2017aep}). The reason why we can have viable DM phenomenology with a fine tuning which is lower than in other models is two-fold: on the one hand, the contribution to
$\hat T$ from the second doublet is positive, so that a
smaller fine tuning is allowed; in addition, because of the richness of the model, it is possible to have a cancellation in the
amplitude relevant for DD, resulting in a viable region of the parameter space compatible with EWPTs.

What happens varying $\beta$ is less trivial: by increasing $\beta$, we increase the effect of subleading terms and the coefficient dependence, and this can in principle affect the results; however, we verified that this is not the case for the values of $\beta$ allowed by EWPTs.

\subsubsection*{Large mass range ($\boldsymbol{m_\eta\approx 400\:\mathrm{GeV}}$)}

The third region of interest is for $m_\eta\approx\SI{400}{\giga\electronvolt}$. As already stated, its existence is due to a cancellation in the main contributions to the relic density between the terms with the exchanges of $h$ and $H_0$. The correct relic abundance can only be reproduced in the loosely natural range of the $c_i$ coefficients. Our benchmark point is already excluded by current DD constraints, and in order to evade these limits one would need $\xi\lesssim 0.01$. For this reason, we do not study this region further.

\section{Non-thermal dark matter production}
\label{sec:non_thermal_DM}

With the larger number of pseudo-Goldstones present in this model, another possible production mechanism for the DM candidate $\eta$, other than the thermal freeze-out discussed above, could be via the decay of heavier pNGBs.

It turns out that the interesting case is when $\kappa$ and $\eta$ are close enough in mass to allow for a sufficiently long decay of $\kappa$ into $\eta$. From~\cref{fig:mass_spectrum}, we see that this happens at roughly $\theta_t\approx\pi/4$. 

In this scenario, the $\eta$ relic density receives two contributions. The first is due to the standard thermal DM freeze-out. The second mechanism is due to the freeze-out of $\kappa$, which is sufficiently long-lived, followed by its decay into $\eta$. Since a single $\eta$ is produced per each decay, the $\kappa$ number-abundance is completely converted into $\eta$. The total $\eta$ relic density is thus given by (see e.g. ref.~\cite{Fairbairn:2008fb}):
\begin{equation}
\Omega_{\rm DM} h^2 = \Omega_\eta h^2 + 
\frac{m_\eta}{m_\kappa} \Omega_\kappa h^2\,.
\end{equation}
It is important to notice that since $H_0$, $A_0$, $H_+$ are all lighter than $\eta$ (cf.~\cref{fig:mass_spectrum}), they also have to be included as final states of $\eta$- and $\kappa$-annihilations. On the other hand, as in standard coannihilations, processes which simply convert $\eta$ in $\kappa$ (and viceversa), are not relevant for the determination of the relic density. We do not report the formulas for all the annihilation channels, but these can be easily computed from the effective Lagrangian given in~\cref{eq:CCWZLagrExpanded} and \cref{eq:NGBfermCoupl,eq:NGBgluonCoupl,eq:NGBNGBCoupl}.

As shown below, in the parameter space of interest the relevant decay channel is only $\kappa\to\eta b\bar b$.
The decay width for the process $\kappa\to\eta q\bar q$ is:
\begin{align}\notag
\Gamma_{\kappa\to\eta q\bar q} &= \frac{3}{32\pi^3m_\kappa}\,\frac{m_q^2}{v^4}|g_{\eta\kappa q}|^2 \int_{m_\eta}^{\frac{m_\kappa^2+m_\eta^2-4m_q^2}{2m_\kappa}}dq_0\sqrt{q_0^2-m_\eta^2}\,(m_\kappa^2+m_\eta^2-2m_\kappa q_0)\\[.2truecm]
&\quad \sqrt{1-\frac{4m_q^2}{m_\kappa^2+m_\eta^2-2m_\kappa q_0}}\,.
\end{align}
The $\kappa$ lifetime has to be compared to the age of the universe at the time of the freeze-out of $\eta$, which is given by:
\begin{equation}
\frac{t_F}{\SI{1}{\second}} \approx \frac{1.5^2}{\sqrt{g_{*,EU}}}{\left(\frac{\SI{1}{\mega\electronvolt}}{m_\eta}\right)}^2x_F^2\,,
\end{equation}
with $g_{*,EU}\approx100$ being the effective number of relativistic species in the early universe and $x_F\approx 25$.

\begin{figure}[t]
\centering
\includegraphics[scale=0.4]{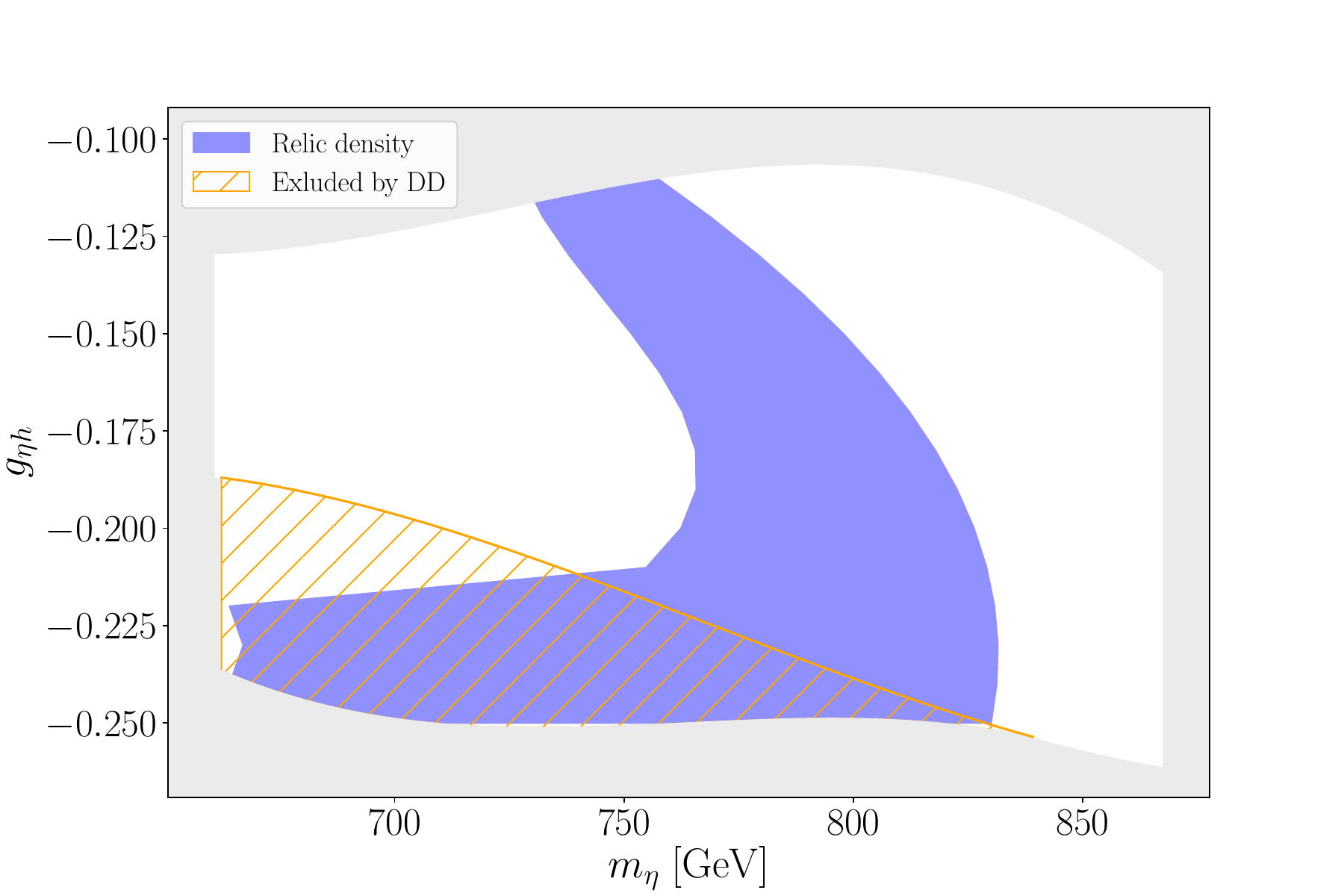}
\caption{\label{fig:non_thermal_relic}Relic density from non-thermal DM production for $\xi=0.01$, $\beta=0.2$, $y_L=1=y_R$, $g_*=3$ and loosely natural $\mathcal O(1)$ coefficients. The blue region corresponds to the $3\sigma$-relic density contour; the orange hatched region is excluded by DD; in the gray region the $\mathcal O(1)$ coefficients are outside the range $[0.1,10]$; finally, ID does not exclude any portion of the parameter space.}
\end{figure}

Given that $\Gamma_{\kappa\to\eta q\bar q}\propto m_q^2$, if $\Delta m_{\kappa,\eta} \equiv m_\kappa-m_\eta>2m_t$, the decay into $t\bar t$ is so quick that $\kappa$ always decays before $\eta$ freezes-out. This can be avoided by either having $\eta$ and $\kappa$ close in mass or by taking $g_{\eta\kappa q}$ very small; we choose the first option as it requires less fine tuning. On the other hand, if $\Delta m_{\kappa,\eta} <2m_t$, the two contributions to the relic density are of the same order, i.e. $\Omega_\eta h^2\approx\Omega_\kappa h^2$. Compared to the thermal case, a higher fine tuning on $\xi$ is needed in order to evade DD constraints. The phenomenology has substantially changed from the thermal case because, by requiring that the two singlets are close in mass, we have put ourselves in the portion of parameter space near $\theta_t=\pi/4$, where both singlets are very heavy.

Analogously to what we did in~\cref{fig:relic_thermal}, we show in~\cref{fig:non_thermal_relic} the results in the $m_\eta$-$g_{\eta h}$ plane, for $\xi=0.01$, $\beta=0.2$, $y_L=1=y_R$, $g_*=3$ and loosely natural coefficients. While DD excludes a portion of the parameter space, current bounds from ID are ineffective. In general, a fine tuning on the masses is needed, since in all the plane $\SI{20}{\giga\electronvolt}\lesssim\Delta m_{\kappa,\eta} \lesssim\SI{50}{\giga\electronvolt}$.

As one can see from~\cref{eq:m_eta}, this range of masses for $m_\eta$ roughly corresponds to $\theta_t\approx\pi/4$, as anticipated. While large mass splittings tend to favour a fast decay for $\kappa$, non-thermal effects are always possible for small mass splittings, although a larger and larger unnaturalness of the coefficients is required (corresponding to a larger and larger fine tuning for $\Delta m_{\kappa,\eta} /m_\eta$).

The non-thermal production mechanism represents an intriguing feature of this model; we think this is one of the most peculiar and interesting aspects of the model. The greater level of complexity with respect to the minimal case has been traded for a richer spectrum of NGBs which can play an active role in DM phenomenology.

\section{Conclusions}
\label{sec:Conclusions}

In this paper we carried out a detailed construction of
the  CH model based on the symmetry breaking pattern
$\SO(7)\to\SO(5)\times\SO(2)$.
The lightest pNGB, $\eta$, is electrically neutral, stable and is a potential DM candidate. 

We studied the DM phenomenology by requiring correct relic abundance
and evading the constraints coming from invisible Higgs decay,  direct detection and indirect detection,
and found large portions of parameter space satisfying all of them 
(see~\cref{sec:ThDM}).
In particular, we identified a viable region of DM mass around
$m_\eta\approx 130\div 160$ GeV (see~\cref{fig:relic_mass_f}),
which is realized for a symmetry breaking scale as low as the minimum
required by the ElectroWeak Precision Tests ($f\gtrsim 0.8$ TeV).
This is mostly due to a cancellation in the couplings of $\eta\eta$ to $q\bar q$
between Higgs-exchange and $H_0$-exchange in the $s$-channel,
which allows to enhance the relic abundance and deplete the direct detection
cross section.
This feature is peculiar of the low-energy 2HDM-like structure of the model.

Another important aspect of the model is that the extra pNGB $\kappa$
may freeze-out in the early universe and subsequently decay to $\eta$, 
thus providing an extra (non-thermal) contribution to the DM density
(see~\cref{sec:non_thermal_DM}).
To the best of our knowledge, no other  model studied so far for 
DM in the CH framework
provides such a possibility.

The main results of this paper may be summarized as follows:
\begin{itemize}
\item the CH model based on the symmetry breaking pattern
$\SO(7)\to\SO(5)\times\SO(2)$  delivers a viable DM candidate, 
consistent with the current phenomenological constraints;
\item the correct amount of DM can be achieved with relatively low 
amount of fine tuning on the symmetry breaking scale $f\gtrsim 0.8$ TeV;
\item it is possible to produce DM also non-thermally via late-time decays
of the heavier pNGB.
\end{itemize}
A more exhaustive study of the indirect detection constraints (e.g. using anti-protons data), of the detection prospects at future colliders, as well
as a detailed analysis of the UV completion of the theory (by including, for instance, the top partners) 
is beyond the scope of the present paper and left for future work.

\section*{Acknowledgements}

We would like to thank Joe Davighi for insightful discussions. DM is grateful to the Mainz Institute for Theoretical Physics (MITP) for its hospitality and its partial support during the completion of part of this work. DM also acknowledges partial support by the INFN grant SESAMO.

\appendix

\section{Generators}
\label{app:generators}

We introduce in this section the generators for the breaking pattern $\SO(7)\to\SO(5)'\times\SO(2)'$; they are defined as: 
\begin{subequations}
\begin{align}
\label{eq:t_L}
{\left(T^\alpha_L\right)}_{ab} &= -\frac{i}{2}\left[\epsilon_{\alpha\beta\gamma}\,\delta_{\beta a}\,\delta_{\gamma b} + \left(\delta_{\alpha a}\,\delta_{4b}-\delta_{\alpha b}\,\delta_{4a}\right)\right]\\[.2truecm]
\label{eq:t_R}
{\left(T^\alpha_R\right)}_{ab} &= -\frac{i}{2}\left[\epsilon_{\alpha\beta\gamma}\,\delta_{\beta a}\,\delta_{\gamma b} - \left(\delta_{\alpha a}\,\delta_{4b}-\delta_{\alpha b}\,\delta_{4a}\right)\right]\\[.2truecm]
\label{eq:t_5}{\left(T^\omega_5\right)}_{ab} &= -\frac{i}{\sqrt2}\left[\delta_{\omega a}\,\delta_{5b} - \delta_{5a}\,\delta_{\omega b}\right]\\[.2truecm]
\label{eq:t_2}
{\left(T_2\right)}_{ab} &= -\frac{i}{\sqrt2}\left[\delta_{6a}\,\delta_{7b} - \delta_{7a}\,\delta_{6b}\right]\\[.2truecm]
\label{eq:t_h1}
{\left(\hat T_1^i\right)}_{ab} &= -\frac{i}{\sqrt2}\left[\delta_{ia}\,\delta_{6b} - \delta_{6a}\,\delta_{ib}\right]\\[.2truecm]
\label{eq:t_h2}
{\left(\hat T_2^i\right)}_{ab} &= -\frac{i}{\sqrt2}\left[\delta_{ia}\,\delta_{7b} - \delta_{7a}\,\delta_{ib}\right]
\end{align}
\end{subequations}
where $\alpha,\beta,\gamma=1,2,3$, $a,b=1,\dots,7$, $\omega=1,\dots,4$ and $i=1,\dots,5$\,. $T_{L,R}$ are the generators of $\SO(4)'\subset\SO(5)'$, $T_5$ are the remaining generators of $\SO(5)'$, and $T_2$ is the generator of $\SO(2)'$; finally, $\hat T_{1,2}$ are the broken generators.

If we define $T^A\equiv\left\{T_L,T_R,T_5\right\}$, we have the following commutation relations:
\begin{align}
\begin{split}
&[\hat T_1^i,\hat T_1^j] = {\left(t^A\right)}_{ij} T^A = [\hat T_2^i,\hat T_2^j] \quad,\quad [\hat T_1^i,\hat T_2^j]=-\frac{i}{\sqrt2}\,\delta_{ij}T_2 \quad,\quad [T^A,T_2]=0\quad,\\[.2truecm]
&[\hat T_{1,2}^i,T^A] = {\left(t^A\right)}_{ij} \hat T^j_{1,2} \quad,\quad [\hat T_1^i,T_2] = \frac{i}{\sqrt2} \hat T_2^i \quad,\quad [\hat T_2^i,T_2] = -\frac{i}{\sqrt2} \hat T_1^i\,,
\end{split}
\end{align}
where $t^A$ is the upper $5\times5$ block of $T^A$, together with:
\begin{align}
\begin{split}
&[T_{L,R}^\alpha,T_{L,R}^\beta] = i\epsilon_{\alpha\beta\gamma}T_{L,R}^\gamma \quad,\quad [T_L^\alpha,T_5^4]=-\frac{i}{2}\,T_5^\alpha \quad,\quad [T_R^\alpha,T_5^4]=\frac{i}{2}\,T_5^\alpha \quad,\\[.2truecm]
&[T_L^\alpha,T_5^\beta]=\frac{i}{2}\left(\delta_{\alpha\beta}T_5^4+\epsilon_{\alpha\beta\gamma}\,T_5^\gamma\right) \quad,\quad [T_R^\alpha,T_5^\beta]=\frac{i}{2}\left(-\delta_{\alpha\beta}T_5^4+\epsilon_{\alpha\beta\gamma}\,T_5^\gamma\right)\quad,\\[.2truecm]
&[T_5^{\omega_1},T_5^{\omega_2}]=\frac{i}{2}\,\epsilon_{\omega_1\omega_2\omega_3}\left(T_L^{\omega_3}+T_R^{\omega_3}\right) \quad,\quad [T_L^\alpha,T_R^\beta]=0\,.
\end{split}
\end{align}

\section{Higgs couplings fit and EWPTs}
\label{app:EWPTHiggs}

To obtain the constraints on the parameters of our model from Higgs measurements we use the recent global Higgs coupling analysis published by ATLAS with 80 fb$^{-1}$ of luminosity \cite{ATLAS:2018doi}.  Restricting the fit to the deviations relevant in the model one has:
\begin{equation}
    \left( \begin{array}{c}
         g_W \\ g_Z \\ k_t \\ k_b
    \end{array}\right) =
    \left( \begin{array}{c}
         1.039 \pm 0.074 \\ 1.067 \pm 0.095 \\ 1.037 \pm 0.088 \\ 1.03 \pm 0.15
    \end{array}\right)~, \qquad
    \rho =  \left( \begin{array}{cccc}
         1 & 0.65 & 0.30 & 0.82 \\
         0.63 & 1 & 0.01 & 0.64 \\
         0.30 & 0.01 & 1 & 0.56 \\
         0.82 & 0.64 & 0.56 & 1
    \end{array}\right)~,
\end{equation}
where $\rho$ is the correlation matrix of the uncertainties on
the  couplings $\{g_W, g_Z, k_t, k_b\}$.
The definitions of $g_{W,Z}$ and $k_{t,b}$ are reported in~\cref{eq:CCWZLagrExpanded,eq:NGBfermCoupl}, while their expressions in terms of the parameters of the model are explicited in~\cref{app:effective_coupl}.

\begin{figure}[t]
\centering
\includegraphics[width=0.47\hsize]{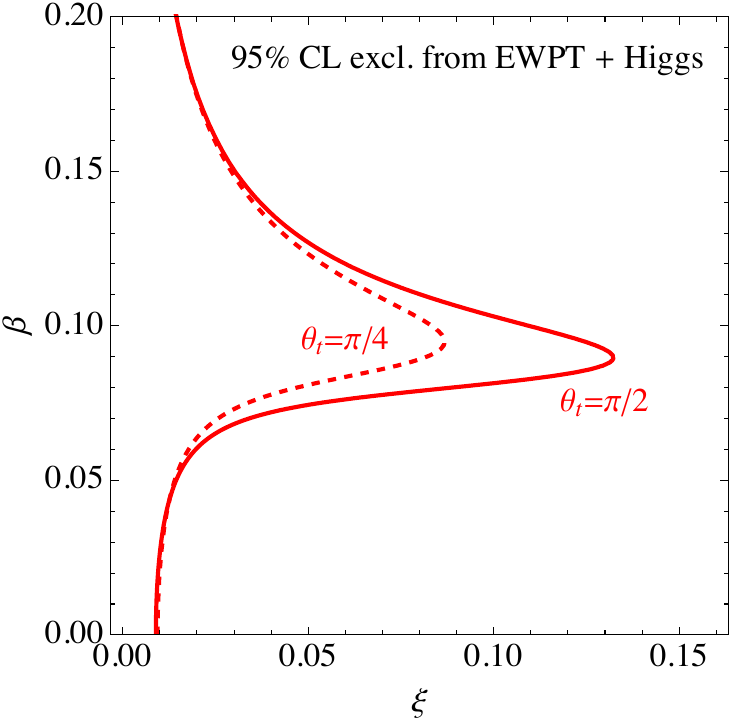}\quad
\includegraphics[width=0.47\hsize]{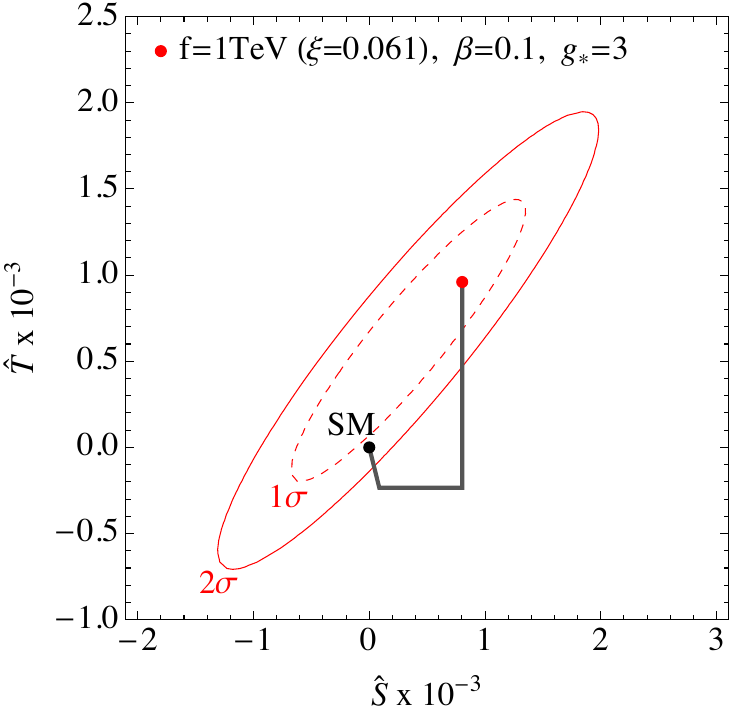}
\caption{\label{fig:HiggsEWPTfit}Left: 95\%CL exclusion limits as a function of $\xi$ and $\beta$ from the combination of Higgs and EWPT data. Solid (dashed) lines are for the $\theta_t = \pi/2 \, (\pi/4)$, while we fixed $\theta_b = 0$. Right: plot of the different contributions to the $\hat{S}$ and $\hat{T}$ parameters in the model for our benchmark point.}
\end{figure}

For the EWPT, we use the updated combined limits on $S = 4 s_W^2 \hat S / \alpha$ and $T = \hat T / \alpha$ from GFitter \cite{Haller:2018nnx}. The dependence of the $\hat S$ and $\hat T$ parameters \cite{Barbieri:2004qk} on the model's parameters is:
\begin{subequations}\begin{align}
    &(\Delta \hat S)_{\rm IR} \approx \frac{g^2}{192 \pi^2} \xi \log \frac{\Lambda^2}{m_h^2}~, \quad
    (\Delta \hat T)_{\rm IR} \approx - \frac{3 g^2}{64 \pi^2} \tan^2 \theta_W \xi \log \frac{\Lambda^2}{m_h^2}~, \\
    &(\Delta \hat S)_{\rm UV} \approx \frac{m_W^2}{\Lambda^2}~, \quad
    (\Delta \hat T)_{\rm UV} = 0~, \\
    &(\Delta \hat S)_{\rm 2HDM} = 0~, \quad
    (\Delta \hat T)_{\rm 2HDM} = \frac{\xi}{4}\left( 1 - \cos 4\beta \right)~, 
\end{align}\end{subequations}
where we use $\Lambda = g_* f$ and $g_*=3$.

In~\cref{fig:HiggsEWPTfit} (left) we show the exclusion limits from the combination of Higgs and EWPT data. Solid (dashed) lines correspond to a value of $\theta_t = \pi/2 \, (\pi/4)$ while we fix $\theta_b = 0$.
The reason why in the combined fit for values $\beta \approx 0.1$ the limit on $\xi$ relaxes up to almost the maximum value allowed by Higgs data is due to the fact that a small but non-zero $\beta$ induces a positive contribution to $\hat T$, which helps to relax the EWPT bound, as is shown in the right panel for the specific benchmark point used throughout the paper.

\section{Expression of the effective couplings}
\label{app:effective_coupl}

In order to describe the effective NGB interaction couplings in terms of the parameters of the potential,
it is convenient to introduce a new set of coefficients. Since each invariant generating the potential brings some power of the fundamental couplings, it is useful to define:
\begin{equation*}
\begin{alignedat}{5}
&\tilde c_y^{(1)} \equiv N_c\, c^{(1)}_{(1,0)}\,y_L^2 &&\quad
&& \tilde c_y^{(2)} \equiv N_c\,c^{(1)}_{(0,1)}\,y_R^2 &&\quad
&&\tilde c_y^{(3)} \equiv N_c\, c^{(1)}_{(2,0)}\,y_L^4/g_*^2 \\
&\tilde c_y^{(4)} \equiv N_c\, c^{(1)}_{(1,1)}\,y_L^2y_R^2/g_*^2 &&\quad
&&\tilde c_y^{(5)} \equiv N_c\, c^{(1)}_{(0,2)}\,y_R^4/g_*^2 &&\quad
&&\tilde c_y^{(6)} \equiv N_c\, c^{(2)}_{(2,0)}\,y_L^4/{(4\pi)}^2 \\
&\tilde c_y^{(7)} \equiv N_c\, c^{(2)}_{(1,1)}\,y_L^2y_R^2/{(4\pi)}^2 &&\quad
&&\tilde c_y^{(8)} \equiv N_c\, c^{(2)}_{(0,2)}\,y_R^4/{(4\pi)}^2 &&\quad
&&\tilde c_y^{(9)} \equiv N_c\, c^{(3)}_{(0,2)}\,y_R^4/{(4\pi)}^2
\end{alignedat}
\end{equation*}
\begin{equation*}
\begin{alignedat}{3}
&\tilde c_g^{(1)}\equiv c_{g'}^{(1)}\,{g'}^2 &&\quad
&&\tilde c_g^{(2)}\equiv c_{g'}^{(2)}\,{g'}^2\\
&\tilde c_g^{(3)}\equiv c_{g}^{(1)}\,g^2 &&\quad
&&\tilde c_g^{(4)}\equiv c_{g'}^{(2)}\,g^2
\end{alignedat}
\end{equation*}
In the following, we provide the list of all the effective couplings used in the computations. The couplings have been listed separately depending on the particles involved in the interaction.

\subsection{Interactions between NGBs and gauge bosons}

The first set of couplings is generated by CCWZ and involves the massive gauge bosons:
\begin{subequations}
\begin{alignat}{3}\label{eq:eff_coefficients_NGB_gauge}
g_V &\approx \sqrt{1 - \xi} , \qquad &&\,\,\,\,b_h &&\approx 1 - 2 \xi, \\[.2truecm]
\lambda_\eta^{(V)} &\approx 2\xi, \qquad &&\lambda_\kappa^{(V)} &&\approx -\xi \beta^2, \\[.2truecm]
g_{H_0W} &\approx - \frac{\beta\xi}{2}, \qquad &&g_{H_0Z} &&\approx \frac{3\beta\xi}{2}, \qquad\quad g_{H_+V} \approx \xi \beta.\\[.2truecm]\notag
\end{alignat}
\end{subequations}
As mentioned in the text, in order to properly discuss LHC phenomenology the couplings of the NGB to the gluons generated by loop of top-quarks has to be included:
\begin{subequations}[resume]
\begin{alignat}{2}
g_{gh} &= -i\,\frac{\alpha_S}{8\pi}\,\tau_h\left[1+(1-\tau_h)f(\tau_h)\right] ,\qquad  &&\tau_h=\frac{4 m_t^2}{m_h^2},\\[.2truecm]
g_{gH_0} &= -i\, k_{H_0t}\frac{\alpha_S}{8\pi}\tau_{H_0}[1+(1-\tau_{H_0})f(\tau_{H_0})],\qquad  &&\tau_{H_0}=\frac{4 m_t^2}{m_{H_0}^2},\\[.2truecm]
g_{g\eta} &= -i\, g_t\frac{\alpha_S}{8\pi}\, \tau_\eta[1+(1-\tau_\eta)f(\tau_\eta)] ,\qquad  &&\tau_{\eta}=\frac{m_t^2}{m_\eta^2}.
\end{alignat}
\end{subequations}
where $f(\tau_X)$ is the usual function appearing for gluon effective couplings (see for instance eq.~(1.198) of~\cite{Plehn:2009nd}).

Notice that since this vertex is generated by a quark-loop the resulting coupling for $H_0$ and $\eta$ will be suppressed by a factor $\xi$ with respect to the one to the SM Higgs and so it is not expected to play a relevant role, although it has been included in our simulations.

\subsection{Interactions between NGBs and fermions}

Interactions with quarks are generated by the partial compositeness Lagrangian and depend on the specific embedding of the fermions:
\begin{subequations}
\begin{align}\label{eq:eff_coefficients_NGB_fermions}
k_q&\approx1-\frac{7}{6}\xi-\frac{\xi}{3}\frac{\cos(3\beta+\alpha_q\theta_q)}{\cos(\beta-\alpha_q\theta_q)}\,,\\[.2truecm]
g_q&\approx-2\xi\frac{\cos\beta\cos\theta_q}{\cos(\beta-\alpha_q\theta_q) }\,,\\[.2truecm]
k_{H_0q}&\approx\frac{2\xi \sin(4\beta)+(-6+\xi)\sin(2\beta-2\alpha_q\theta_q)+4\xi\sin(2\beta+2\alpha_q\theta_q)}{12\cos^2(\beta-\alpha_q\theta_q)}\,,\\[.2truecm]
g_{\kappa q} &\approx-2\alpha_q\xi\frac{\sin\beta\sin\theta_q}{\cos(\beta-\alpha_q\theta_q)}\,,\\[.2truecm]
g_{\eta\kappa q} &\approx-i\alpha_q\xi \tan(\beta-\alpha_q\theta_q)\,.
\end{align}
\end{subequations}
with $\alpha_q=1$ ($-1$) for quarks with charge $2/3\;(-1/3)$.

\subsection{Interactions among NGBs}

Since the potential has been generated using an expansion of the NGB matrix, we list here only the relevant orders in $\xi$ and $\beta$:
\begin{subequations}
\begin{align}\notag
\label{eq:g_eta}
g_{\eta h} \approx& -  \frac{g_*^2}{8\pi^2}\,\cos^2\theta_t\left[2\tilde c^{(5)}_y-\tilde c^{(7)}_y+2\tilde c^{(8)}_y+2\cos(2\theta_t)(\tilde c^{(5)}_y+\tilde c^{(8)}_y)\right] \\[.2truecm]\notag
& + \frac{g_*^2\beta}{4\pi^2}\cot\theta_t\cos(2\theta_t)(\tilde c^{(2)}_y+2\tilde c^{(7)}_y)\\[.2truecm]
& - \frac{g_*^2\beta^2}{8\pi^2}\cot^2\theta_t\left[\tilde c^{(5)}_y-2\tilde c^{(8)}_y-2\cos(2\theta_t)(\tilde c^{(2)}_y+\tilde c^{(5)}_y+2\tilde c^{(7)}_y-\tilde c^{(8)}_y)+\tilde c^{(5)}_y\cos(4\theta_t)\right]
\\[.2truecm]\notag
\lambda_{\eta h} \approx &\, \frac{g_*^2}{24\pi^2}\,\cos^2\theta_t\left[2\tilde c^{(2)}_y-6\tilde c^{(5)}_y+7\tilde c^{(7)}_y-6\tilde c^{(8)}_y-6\cos(2\theta_t)(\tilde c^{(5)}_y+\tilde c^{(8)}_y)\right] \\[.2truecm]\notag
& + \frac{g_*^2\beta}{4\pi^2}\cot\theta_t\cos(2\theta_t)(\tilde c^{(2)}_y+2\tilde c^{(7)}_y)\\[.2truecm]\notag
& - \frac{g_*^2\beta^2}{24\pi^2}\cot^2\theta_t\big[\tilde c^{(2)}_y+3\tilde c^{(5)}_y+2\tilde c^{(7)}_y-6\tilde c^{(8)}_y-\cos(2\theta_t)(7\tilde c^{(2)}_y+6\tilde c^{(5)}_y+14\tilde c^{(7)}_y-6\tilde c^{(8)}_y) \\[.2truecm]
&\qquad\qquad\qquad+3\tilde c^{(5)}_y\cos(4\theta_t)\big]
\\[.2truecm]\notag
\lambda \approx& \,1-\beta^2 \\[.2truecm]\notag
& -  \frac{g_*^2\beta^2}{16\pi^2}\frac{v^2}{m_h^2}\left[\tilde c_g^{(1)}+\tilde c_g^{(2)}+\tilde c_g^{(3)}+\tilde c_g^{(4)}-16\tilde c^{(2)}_y+\tilde c^{(5)}_y-32\tilde c^{(7)}_y+2\tilde c^{(8)}_y\right.\\[.2truecm]\notag
&\qquad\qquad\quad\, -  \cos(2\theta_t)(14\tilde c^{(2)}_y-2\tilde c^{(5)}_y+29\tilde c^{(7)}_y-2\tilde c^{(8)}_y)\\[.2truecm]
&\qquad\qquad\quad\,\left.+8\csc^2\theta_t(\tilde c^{(2)}_y+2\tilde c^{(7)}_y)+\tilde c^{(5)}_y\cos(4\theta_t)\right] \\[.2truecm]\notag
g_{H_0} \approx&\, \frac{m_h^2}{v^2}\left(1-\beta^2\right)\\[.2truecm]\notag
&- \frac{g_*^2}{8\pi^2}\big[\tilde c_g^{(1)}+\tilde c_g^{(2)}+\tilde c_g^{(3)}+\tilde c_g^{(4)}+\tilde c^{(5)}_y+2\tilde c^{(8)}_y+\cos(2\theta_t)(2\tilde c^{(5)}_y-\tilde c^{(7)}_y+2\tilde c^{(8)}_y)
\notag \\[.2truecm]
&\quad\qquad +\tilde c^{(5)}_y\cos(4\theta_t)\big]\notag\\[.2truecm]\notag
&+ \frac{g_*^2\beta}{4\pi^2}\csc\theta_t[\cos\theta_t+\cos(3\theta_t)](\tilde c^{(2)}_y+2\tilde c^{(7)}_y)\\[.2truecm]\notag
& + \frac{g_*^2\beta^2}{16\pi^2} \bigg[15(\tilde c_g^{(1)}+\tilde c_g^{(2)}+\tilde c_g^{(3)}+\tilde c_g^{(4)}+\tilde c^{(5)}_y+2\tilde c^{(8)}_y) \\[.2truecm]
&\qquad\qquad+ \cos(2\theta_t)(14\tilde c^{(2)}_y+6\tilde c^{(5)}_y+25\tilde c^(_y6\tilde c^{(8)}_y) + 15 \tilde c^{(5)}_y \cos(4\theta_t)\bigg] \\[.2truecm]\notag
g_{H_0hh} \approx &\,\frac{g_*^2\beta}{8\pi^2}\bigg[ 3\left[\tilde c_g^{(1)}+\tilde c_g^{(2)}+\tilde c_g^{(3)}+\tilde c_g^{(4)}+\tilde c^{(5)}_y+2\tilde c^{(8)}_y\right]+\cos(2\theta_t)(4\tilde c^{(2)}_y+6\tilde c^{(5)}_y+5\tilde c^{(7)}_y+6\tilde c^{(8)}_y) \\[.2truecm]\notag
&\qquad\, +3\tilde c^{(5)}_y\cos(4\theta_t)\bigg]\\[.2truecm]
&- \frac{g_*^2\beta^2}{2\pi^2}\cot(2\theta_t)[\cos\theta_t+\cos(3\theta_t)](\tilde c^{(2)}_y+2\tilde c^{(7)}_y) \\[.2truecm]\notag
\lambda_{\eta H_0} \approx & -\frac{g_*^2}{8\pi^2}\cos^2\theta_t\left[2\tilde c^{(5)}_y-\tilde c^{(7)}_y-2\tilde c^{(8)}_y-2\cos(2\theta_t)(\tilde c^{(5)}_y-\tilde c^{(8)}_y)\right] \\[.2truecm]\notag
& + \frac{g_*^2\beta^2}{24\pi^2}\bigg[\tilde c^{(2)}_y-3\tilde c^{(5)}_y+2\tilde c^{(7)}_y-6\tilde c^{(8)}_y - \cos(2\theta_t)\left(7\tilde c^{(2)}_y+6\tilde c^{(5)}_y+14\tilde c^{(7)}_y+6\tilde c^{(8)}_y\right)\\[.2truecm]
&\qquad\qquad- 3\tilde c^{(5)}_y\cos(4\theta_t)\bigg]\\[.2truecm]\notag
\lambda_{H_0} \approx& -\frac{3g_*^2\beta}{8\pi^2}\bigg[\tilde c_g^{(1)}+\tilde c_g^{(2)}+\tilde c_g^{(3)}+\tilde c_g^{(4)}+\tilde c^{(5)}_y+2\tilde c^{(8)}_y-\cos(2\theta_t)(2\tilde c^{(5)}_y-\tilde c^{(7)}_y+2\tilde c^{(8)}_y) +\tilde c^{(5)}_y\cos(4\theta_t)\bigg]\\[.2truecm]
& + \frac{3g_*^2\beta^2}{2\pi^2}\tan\theta_t\cos(2\theta_t)(\tilde c^{(2)}_y+2\tilde c^{(7)}_y) \\[.2truecm]\notag
g_{\eta H_0} \approx& \,\frac{g_*^2\beta}{8\pi^2}\bigg[\tilde c^{(5)}_y+2\tilde c^{(8)}_y+2\cos(2\theta_t)(\tilde c^{(2)}_y+\tilde c^{(5)}_y+2\tilde c^{(7)}_y+\tilde c^{(8)}_y) +\tilde c^{(5)}_y\cos(4\theta_t)\bigg]\\[.2truecm]
&  - \frac{g_*^2\beta^2}{4\pi^2}\tan\theta_t\cos(2\theta_t)(\tilde c^{(2)}_y+2\tilde c^{(7)}_y) \\[.2truecm]\notag
g_{A_0 h} \approx&\frac{g_*^2}{8\pi^2}\left[-\frac{8\pi^2}{g_*^2}\frac{m_h^2}{v^2}+\tilde c^{(5)}_y+\tilde c^{(8)}_y+\cos(2\theta_t)(2\tilde c^{(5)}_y-\tilde c^{(7)}_y+2\tilde c^{(8)}_y)+\cos(4\theta_t)(\tilde c^{(5)}_y+\tilde c^{(8)}_y)\right]\\[.2truecm]\notag
& - \frac{g_*^2\beta}{4\pi^2}\csc\theta_t\sec\theta_t(\tilde c^{(2)}_y+2\tilde c^{(7)}_y)\cos^2(2\theta_t)\\[.2truecm]\notag
& - \frac{g_*^2\beta^2}{16\pi^2}\bigg[-\frac{16\pi^2}{g_*^2}\frac{m_h^2}{v^2}+7\tilde c_g^{(1)}+7\tilde c_g^{(2)}+7\tilde c_g^{(3)}+7\tilde c_g^{(4)}+16\tilde c^{(2)}_y+7\tilde c^{(5)}_y+32\tilde c^{(7)}_y+14\tilde c^{(8)}_y\\[.2truecm]\notag
&\qquad\qquad + \cos(2\theta_t)(14\tilde c^{(2)}_y+6\tilde c^{(5)}_y+25\tilde c^{(7)}_y+6\tilde c^{(8)}_y)\\[.2truecm]
&\qquad\qquad- 16\csc^2(2\theta_t)(\tilde c^{(2)}_y+2\tilde c^{(7)}_y) + 7\tilde c^{(5)}_y\cos(4\theta_t)\bigg]
\\[.2truecm]\notag
g_{A_0H_0} \approx& -\frac{g_*^2\beta}{8\pi^2}\bigg[\tilde c_g^{(1)}+\tilde c_g^{(2)}+\tilde c_g^{(3)}+\tilde c_g^{(4)}+\tilde c^{(5)}_y+2\tilde c^{(8)}_y-\cos(2\theta_t)(2\tilde c^{(5)}_y-\tilde c^{(7)}_y+2\tilde c^{(8)}_y)+\tilde c^{(5)}_y\cos(4\theta_t)\bigg]\\[.2truecm]
& + \frac{g_*^2\beta^2}{2\pi^2}\tan\theta_t\cos(2\theta_t)(\tilde c^{(2)}_y+2\tilde c^{(7)}_y) \\[.2truecm]\notag
g_{\kappa h} \approx&\, \frac{g_*^2}{8\pi^2}\sin^2\theta_t\left[\tilde c^{(7)}_y-4(\tilde c^{(5)}_y-\tilde c^{(8)}_y)\cos^2\theta_t\right] \\[.2truecm]
& - \frac{g_*^2\beta^2}{8\pi^2}\left[\tilde c^{(5)}_y+2\tilde c^{(8)}_y-2\cos(2\theta_t)(\tilde c^{(2)}_y+\tilde c^{(5)}_y+2\tilde c^{(7)}_y+\tilde c^{(8)}_y)+\tilde c^{(5)}_y\cos(4\theta_t)\right]\\[.2truecm]
g_{\kappa H_0} \approx& -\frac{g_*^2\beta}{2\pi^2}\sin^2\theta_t(\tilde c^{(8)}_y-\tilde c^{(5)}_y\cos(2\theta_t))+ \frac{g_*^2\beta^2}{4\pi^2}\tan\theta_t\cos(2\theta_t)(\tilde c^{(2)}_y+2\tilde c^{(7)}_y) \\[.2truecm]\notag
\lambda_{\kappa h} \approx&\, \frac{g_*^2}{8\pi^2}\sin^2\theta_t(\tilde c^{(7)}_y-4\cos^2\theta_t(\tilde c^{(5)}_y-\tilde c^{(8)}_y))\\[.2truecm]\notag
& + \frac{g_*^2\beta^2}{24\pi^2}\bigg[\tilde c^{(2)}_y-3\tilde c^{(5)}_y+2\tilde c^{(7)}_y-6\tilde c^{(8)}_y + \cos(2\theta_t)(5\tilde c^{(2)}_y+6\tilde c^{(5)}_y+10\tilde c^{(7)}_y+6\tilde c^{(8)}_y) \\[.2truecm]
&\qquad\qquad- 3\tilde c^{(5)}_y\cos(4\theta_t)\bigg] \\[.2truecm]\notag
\lambda_{\kappa H_0} \approx &  -\frac{g_*^2}{24\pi^2}\bigg[2\left(2 \tilde{c}_y^{(7)}+\tilde{c}_y^{(2)}\right)\cos^2\theta_t-\sin^2\theta_t\left(4\tilde{c}_y^{(2)}+11 \tilde{c}_y^{(7)}\right)+12\sin^4\theta _t\left(\tilde{c}_y^{(5)}+ \tilde{c}_y^{(8)}\right)\bigg]\\[.2truecm]\notag
& +\frac{g_*^2\beta}{24\pi^2}[3-5\cos(2\theta_t)]\csc\theta_t\sec\theta_t\cos(2\theta_t)(\tilde c^{(2)}_y+2\tilde c^{(7)}_y)\\[.2truecm]\notag
&-\frac{g_*^2\beta^2}{24\pi^2}\bigg[13\tilde c^{(2)}_y-3\tilde c^{(5)}_y+26\tilde c^{(7)}_y-6\tilde c^{(8)}_y-\cos(2\theta_t)(13\tilde c^{(2)}_y-3\tilde c^{(5)}_y+26\tilde c^{(7)}_y-6\tilde c^{(8)}_y)\\[.2truecm]
& \qquad\qquad+2 (\csc^2\theta_t-4\sec^2\theta_t)(\tilde c^{(2)}_y+2\tilde c^{(7)}_y)\bigg] \\[.2truecm]\notag
\lambda_{\eta A_0} \approx &\, \frac{g_*^2}{8\pi^2}\,\cos^2\theta_t \left[2\tilde{c}_y^{(5)}-\tilde{c}_y^{(7)}-2\tilde{c}_y^{(8)}-2\cos(2\theta_t)(\tilde{c}_y^{(5)}-\tilde{c}_y^{(8)})\right] \\[.2truecm]\notag
& - \frac{\beta^2g_*^2}{24\pi^2} \left[\tilde{c}_y^{(2)}-3\tilde{c}_y^{(5)}+2\tilde{c}_y^{(7)}-6\tilde{c}_y^{(8)}-\cos(2\theta_t)(7\tilde{c}_y^{(2)}+6\tilde{c}_y^{(5)}+14\tilde{c}_y^{(7)}+6\tilde{c}_y^{(8)})\right.\\[.2truecm]
&\left.\qquad\qquad-3\tilde{c}_y^{(5)}\cos(4\theta_t)\right] \\[.2truecm]\notag
\lambda_{\eta H_+} \approx &\, \frac{g_*^2}{8\pi^2}\,\cos^2\theta_t \left[2\tilde{c}_y^{(5)}-\tilde{c}_y^{(7)}-2\tilde{c}_y^{(8)}-2\cos(2\theta_t)(\tilde{c}_y^{(5)}-\tilde{c}_y^{(8)})\right] \\[.2truecm]\notag
& - \frac{\beta^2g_*^2}{24\pi^2} \left[\tilde{c}_y^{(2)}-3\tilde{c}_y^{(5)}+2\tilde{c}_y^{(7)}-6\tilde{c}_y^{(8)}-\cos(2\theta_t)(7\tilde{c}_y^{(2)}+6\tilde{c}_y^{(5)}+14\tilde{c}_y^{(7)}+6\tilde{c}_y^{(8)})\right.\\[.2truecm]
&\left.\qquad\qquad-3\tilde{c}_y^{(5)}\cos(4\theta_t)\right] \\[.2truecm]\notag
g_{H_+h} \approx & -\frac{g_*^2}{16\pi^2}\, \left[\tilde{c}_y^{(5)}+2\tilde{c}_y^{(6)}-2\tilde{c}_y^{(7)}-\tilde{c}_y^{(8)}-\cos(4\theta_t)(\tilde{c}_y^{(5)}-\tilde{c}_y^{(8)})-2\tilde{c}_g^{(3)}-2\tilde{c}_g^{(4)}\right] \\[.2truecm]
& - \frac{\beta^2g_*^2}{4\pi^2} \left[\tilde{c}_y^{(5)}+2\tilde{c}_y^{(8)}+\tilde{c}_y^{(5)}\cos(4\theta_t)+\tilde{c}_g^{(1)}+\tilde{c}_g^{(2)}+\tilde{c}_g^{(3)}+\tilde{c}_g^{(4)}\right] \\[.2truecm]\notag
g_{H_+H_0} \approx & - \frac{\beta g_*^2}{8\pi^2}\,\cos^2\theta_t \bigg[\tilde{c}_y^{(5)}+2\tilde{c}_y^{(8)}-\cos(2\theta_t)(2\tilde{c}_y^{(5)}-\tilde{c}_y^{(7)}+2\tilde{c}_y^{(8)})+\tilde{c}_y^{(5)}\cos(4\theta_t) \\[.2truecm]\notag
& \qquad\qquad\qquad+\tilde{c}_g^{(1)}+\tilde{c}_g^{(2)}+\tilde{c}_g^{(3)}+\tilde{c}_g^{(4)}\bigg]\\[.2truecm]
&+\frac{\beta^2g_*^2}{2\pi^2}(\tilde{c}_y^{(2)}+2\tilde{c}_y^{(7)})\cos(2\theta_t)\tan\theta_t \\[.2truecm]\notag
\lambda_{\kappa A_0} \approx&\, \frac{g_*^2}{24\pi^2} \left[2(\tilde{c}_y^{(2)}+2\tilde{c}_y^{(7)})\cos^2\theta_t - \sin^2\theta_t [4\tilde{c}_y^{(2)}+11\tilde{c}_y^{(7)}-12\sin^2\theta_t(\tilde{c}_y^{(5)}+\tilde{c}_y^{(8)})]\right]\\[.2truecm]\notag
& + \frac{\beta g_*^2}{24\pi^2} (\tilde{c}_y^{(2)}+2\tilde{c}_y^{(7)})\cos(2\theta_t)[5\cos(2\theta_t)-3]\csc\theta_t\sec\theta_t\\[.2truecm]\notag
& + \frac{\beta^2g_*^2}{24\pi^2}\bigg[13\tilde{c}_y^{(2)}-3\tilde{c}_y^{(5)}+26\tilde{c}_y^{(7)}-6\tilde{c}_y^{(8)}-\cos(2\theta_t)(13\tilde{c}_y^{(2)}-6\tilde{c}_y^{(5)}+26\tilde{c}_y^{(7)}-6\tilde{c}_y^{(8)}) \\[.2truecm]
& \qquad\qquad-3\tilde{c}_y^{(5)}\cos(4\theta_t)+2(\tilde{c}_y^{(2)}+2\tilde{c}_y^{(7)})(\csc^2\theta_t-4\sec^2\theta_t)\bigg]\\[.2truecm]\notag
\lambda_{\kappa H_+} \approx& \,\frac{g_*^2}{24\pi^2} \left[2(\tilde{c}_y^{(2)}+2\tilde{c}_y^{(7)})\cos^2\theta_t - \sin^2\theta_t [4\tilde{c}_y^{(2)}+11\tilde{c}_y^{(7)}-12\sin^2\theta_t(\tilde{c}_y^{(5)}+\tilde{c}_y^{(8)})]\right]\\[.2truecm]\notag
& + \frac{\beta g_*^2}{24\pi^2} (\tilde{c}_y^{(2)}+2\tilde{c}_y^{(7)})\cos(2\theta_t)[5\cos(2\theta_t)-3]\csc\theta_t\sec\theta_t\\[.2truecm]\notag
& + \frac{\beta^2g_*^2}{24\pi^2}\bigg[13\tilde{c}_y^{(2)}-3\tilde{c}_y^{(5)}+26\tilde{c}_y^{(7)}-6\tilde{c}_y^{(8)}-\cos(2\theta_t)(13\tilde{c}_y^{(2)}-6\tilde{c}_y^{(5)}+26\tilde{c}_y^{(7)}-6\tilde{c}_y^{(8)}) \\[.2truecm]
&\qquad\qquad -3\tilde{c}_y^{(5)}\cos(4\theta_t)+2(\tilde{c}_y^{(2)}+2\tilde{c}_y^{(7)})(\csc^2\theta_t-4\sec^2\theta_t)\bigg]\\[.2truecm]
k_\mathrm{der} = & \frac{2\xi}{3}
\end{align}
\end{subequations}


\section{Relic density}
\label{app:relicdensity}

The main contributions to the relic density are given by $\eta$ annihilations into $h$, $W$, $Z$, $t$ and $b$; below the $W,\,Z$ threshold, however, also the channels $\eta\eta\to WW^*,\,ZZ^*$ have to be included. With the vertices given in \cref{app:effective_coupl}, the thermally-averaged cross sections are:
\begin{subequations}
\begin{align}\notag
\langle\sigma v_\mathrm{rel}\rangle_{\eta\eta\to hh} &= \frac{1}{64\pi m_\eta^2}\bigg|\lambda_{\eta h}+\frac{3g_{\eta h}\lambda\, m_h^2}{4m_\eta^2-m_h^2+im_h\Gamma_h}-\frac{4g_{\eta H_0}g_{H_0hh}\, v^2}{4m_\eta^2-m_{H_0}^2+i m_{H_0}\Gamma_{H_0}}-\frac{2g_{\eta h}^2v^2}{m_h^2-2m_\eta^2}\\[.2truecm]
&\quad\quad\quad\quad\,\,\,+{\frac{k_\mathrm{der}(5m_\eta^2-m_h^2)}{v^2}\bigg|}^2\sqrt{1-\frac{m_h^2}{m_\eta^2}}\\[.2truecm]\notag
\langle\sigma v_\mathrm{rel}\rangle_{\eta\eta\to VV} &= \frac{\alpha_V}{32\pi m_\eta^2}\frac{m_V^4}{v^4}\left|\lambda_\eta^{(V)}+\frac{2g_{\eta h} g_V\,v^2}{4m_\eta^2-m_h^2+im_h\Gamma_h}-\frac{2g_{\eta H_0} g_{H_0V}\,v^2}{4m_\eta^2-m_{H_0}^2}\right|^2\\[.2truecm]
&\quad\left[2+{\left(\frac{2m_\eta^2-m_V^2}{m_V^2}\right)}^2\right]\sqrt{1-\frac{m_V^2}{m_\eta^2}}\\[.2truecm]
\label{eq:ann_into_top}
\langle\sigma v_\mathrm{rel}\rangle_{\eta\eta\to q\bar q} &= \frac{3}{4\pi}\frac{m_q^2}{v^4}\left|g_q+\frac{g_{\eta h} k_q\,v^2}{4m_\eta^2-m_h^2+im_h\Gamma_h}-\frac{g_{\eta H_0} k_{H_0q}\,v^2}{4m_\eta^2-m_{H_0}^2}\right|^2{\left(1-\frac{m_q^2}{m_\eta^2}\right)}^\frac{3}{2}\,,
\end{align}
\end{subequations}
with $\alpha_V = 1\:(1/2)$ for $W\:(Z)$.

Finally, also the process $\eta\eta\to VV^*$ can play an important role below the $W$ and $Z$ bosons production threshold. The thermally-averaged cross section for this process is (in this case, the exchange of $H_0$ in the $s$-channel is completely negligible):
\begin{equation}
\langle\sigma v_\mathrm{rel}\rangle_{\eta\eta\to VV^*} = \sum_f\frac{k_{(V)}^2N_c^{(f)}}{1536\pi^3 m_\eta^2}\,\frac{m_V^4}{v^4}\left(\lambda_\eta^{(V)}+\frac{2g_{\eta h} g_V\,v^2}{4m_\eta^2-m_h^2}\right)^2 F(\varepsilon_V,\zeta_f)\,,
\end{equation}
where $N_c^{(f)}$ is the number of colors of the final state $f$, and:
\begin{align}\notag
F(\varepsilon_V,\zeta_f) &= \int_{\varepsilon_V}^{1+\frac{\varepsilon_V^2}{4}-\zeta_f^2}dy\,\frac{\sqrt{y^2-\varepsilon_V^2}}{(1-y)^2}\,\frac{1}{\varepsilon_V^2}\sqrt{1-\frac{4\zeta_f^2}{4-4y+\varepsilon_V^2}}\\[.2truecm]\notag
&\quad\bigg\{\left(\tau_{(V)}^2+
\chi_{(V)}^2\right)\left[4y^2-12\varepsilon_V^2y+8\varepsilon_V^2+3\varepsilon_V^4\right]\\[.2truecm]\notag
&\quad +\frac{ 2\zeta_f^2}{4-4y+\varepsilon_V^2}\,\bigg[\tau_{(V)}^2\left(4y^2-12\varepsilon_V^2y+8\varepsilon_V^2+3\varepsilon_V^4\right)\\[.2truecm]
&\quad +2\chi_{(V)}^2\left(2y^2+12\varepsilon_V^2y-14\varepsilon_V^2-3\varepsilon_V^4\right)\bigg]\bigg\}
\end{align}
with:
\begin{equation}
k_{(V)}=
\begin{cases}
\frac{g}{2\sqrt2}\\[.2truecm]
\frac{\sqrt{g^2+{g'}^2}}{2}
\end{cases},\quad \tau_{(V)}=
\begin{cases}
1\\
c_V
\end{cases},\quad \chi_{(V)}=
\begin{cases}
1\quad,\quad \;\;V=W\\
c_A\quad,\quad V=Z
\end{cases}\,,
\end{equation}
$\varepsilon_V\equiv\frac{m_V}{m_\eta}$
and $\zeta_f\equiv(m_{f_1}+m_{f_2})/(2m_\eta)$, $f_1$ and $f_2$ being the final states of $V^*$ decay. Obviously, also the coefficients $\tau_{(V)}$ and $\chi_{(V)}$ depend on the final states.

The relic density is computed from the effective cross section as:
\begin{equation}
\Omega h^2=\frac{0.03}{\displaystyle\int_{x_F}^\infty dx\,\frac{\sqrt{g_*}\,}{x^2}\,\frac{{\langle\sigma v\rangle}_\mathrm{eff}}{\SI{1}{\pico\barn}}}\,,
\label{eq:relic_abundance}
\end{equation}
with:
\begin{equation}
x_F=25+\log\left[\frac{1.67}{\sqrt{g_*x_F}}\,\frac{m_1}{\SI{100}{\giga\electronvolt}}\,\frac{{\langle\sigma v\rangle}_\mathrm{eff}}{\SI{1}{\pico\barn}}\right]\,.
\end{equation}
The same formulas also apply for the case of the non-thermal contributions of~\cref{sec:non_thermal_DM}, where also $\kappa$ freezes-out before decaying into $\eta$.

\bibliographystyle{JHEP}
\bibliography{bibliography}

\end{document}